\documentclass{aastex}
\usepackage{epsfig}
\def\ter{\theta_{\rm er}}
\def\si{\sin \theta}
\def\Ro{R_{\odot}}

\def\Bm{B_{\rm max}}
\newcommand{\Bf}{{\bf B}}
\newcommand{\ep}{{\bf e}_\phi}
\newcommand{\vf}{{\bf v}}
\newcommand{\pa}{\partial}
\newcommand{\Rs}{R_{\odot}}
\newcommand{\er}{\mbox{erf}}
\newcommand{\Bc}{B_c}

\shorttitle{Mean-field formulation of Babcock--Leighton dynamo}

\shortauthors{Nandy \& Choudhuri }                  

\begin{document}

\title{TOWARDS A MEAN-FIELD FORMULATION OF THE BABCOCK--LEIGHTON TYPE
SOLAR DYNAMO.  I.~{\Large $\alpha$}-COEFFICIENT VERSUS DURNEY'S DOUBLE RING
APPROACH} 

\author{Dibyendu Nandy and Arnab Rai Choudhuri}
\affil{Department of Physics, Indian Institute of Science, Bangalore 560012\\
(email: dandy@physics.iisc.ernet.in, arnab@physics.iisc.ernet.in )}
\email{dandy@physics.iisc.ernet.in,arnab@physics.iisc.ernet.in }
 

\begin{abstract}

We develop a model of the solar dynamo in which,
on the one hand, we follow the Babcock--Leighton approach
to include surface processes like the production of poloidal 
field from the decay of active regions, and, on the other hand, 
we attempt to develop a mean field theory that can be studied 
in quantitative detail. One of the  main challenges in 
developing such models is to treat the buoyant rise of toroidal
field and the production of poloidal field from it near the 
surface. The previous paper by Choudhuri, Sch\"ussler, \& Dikpati 
(1995) did not incorporate buoyancy. We extend this model
by two contrasting methods. In one method, we incorporate 
the generation of the poloidal field near the solar surface by 
Durney's procedure of double ring eruption. In the second method, 
the poloidal field generation is treated by a positive 
$\alpha$-effect concentrated near the solar surface, coupled 
with an algorithm for handling buoyancy. The two methods are 
found to give qualitatively similar results.

\end{abstract}

\section{Introduction}
Historically there have been two theoretical approaches in
understanding the origin of the solar magnetic cycle: the
Parker--Steenbeck--Krause--R\"adler (PSKR) approach
(Parker 1955; Steenbeck, Krause \& R\"adler 1966) and the
Babcock--Leighton (BL) approach (Babcock 1961; Leighton 1969).
In both these approaches, the toroidal component of the magnetic
field is supposed to be generated from the poloidal component
by the stretching of field lines due to differential rotation.
In order for a self-sustaining dynamo to exist, the poloidal
field has to be generated back from the toroidal field.  The two
approaches mentioned above attempt to solve this problem in
two different ways.  In the PSKR approach, the cyclonic
turbulence in the interior of the Sun is supposed to twist
the toroidal field lines to produce the poloidal field
(the so-called $\alpha$-effect).  On
the other hand, the BL approach puts more stress on what is
happening at the solar surface and assumes that the poloidal field
arises out of the decay of tilted bipolar active regions 
(which result from the magnetic buoyancy of the toroidal field).
Various aspects of the generation of poloidal field at the surface
have been discussed by Wang \& Sheeley (1991) and Durney, 
De Young, \& Roxburgh (1993).

A formal mathematical formulation of the PSKR approach was
developed on the basis of the mean field magnetohydrodynamics
(Steenbeck, Krause, \& R\"adler 1966; Moffatt 1978, Chap.\ 7;
Parker 1979, \S18.3; Choudhuri 1998, \S16.5).  In 
comparison, the BL approach
was based on rather heuristic, and often 
qualitative, arguments. Until recently,
most of the detailed mathematical models of the solar dynamo
were worked out on the basis of the PSKR approach. Only in the last
few years there have at last been attempts of putting the BL approach
on a mathematical footing comparable in sophistication to
the mathematical theory of the PSKR approach (Choudhuri et
al.\ 1995; Durney 1995, 1996, 1997;
Dikpati \& Charbonneau 1999).  It now appears that the most
successful model of the solar cycle will be something which
incorporates the best features of both these approaches
(Choudhuri 1999).

Since  magnetic buoyancy would be particularly destabilizing
in the main body of the convection zone (Parker 1975;
Moreno-Insertis 1983), several theorists
(Spiegel \& Weiss 1980; van Ballegooijen 1982; DeLuca \&
Gilman 1986; Choudhuri 1990) argued that the solar dynamo may
be operating in the overshoot layer at the bottom of the
convection zone. With the helioseismic discovery of a shear
layer at the bottom of the convection zone, it now appears
fairly certain that the generation of the strong toroidal field
by the stretching of field lines must be taking place there.
However, it seems unlikely that the whole dynamo process (as
envisaged in the PSKR approach) occurs at the bottom of
the convection zone.  The studies of
buoyant rise of the toroidal flux from there suggest that the
toroidal field at the bottom of the convection zone should
be of the order $10^5$ G, substantially stronger than the
equipartition value (Choudhuri \& Gilman 1987; Choudhuri 1989;
D'Silva \& Choudhuri 1993; Fan, Fisher, \& DeLuca 1993; 
Caligari, Moreno-Insertis, \& Sch\"ussler 1995).
Such a strong field would completely quench the $\alpha$-effect
of the PSKR approach.  To explain the generation of the poloidal
field, the most natural way is to invoke the BL idea of the
decay of tilted active regions, though there are still some
attempts to work within the PSKR approach by considering an
interface dynamo (Parker 1993; Charbonneau \& MacGregor 1997;
Markiel \& Thomas 1999).
In this paper, we assume that the poloidal field is produced
by the decay of tilted active regions near the solar surface.

Although the sunspots migrate equatorward with the solar cycle,
the weak diffuse magnetic field on the solar surface migrates
poleward (Bumba \& Howard 1965; Howard \& LaBonte 1981;
Makarov, Fatianov, \& Sivaraman 1983; Makarov \& Sivaraman 1989).  
Most of the dynamo models based on the PSKR approach (starting from
Steenbeck \& Krause 1969) mainly concentrated on the sunspots
and ignored the poleward migration of the weak diffuse field.
The poleward migration has been explained by assuming that
the weak diffuse field (which is essentially the poloidal
field) is carried by the meridional circulation (Wang, Nash, \& 
Sheeley 1989a, 1989b; Dikpati \& Choudhuri 1994, 1995; Choudhuri \& 
Dikpati 1999).  If we now accept the BL idea that the poloidal
field is produced by the decay of tilted bipolar active
regions, then the meridional circulation should play an
important role in the dynamo problem by bringing the
poloidal field from the surface to the bottom of the convection
zone, where the poloidal field is stretched out to produce
the toroidal field. The challenge before us now is to develop
a new type of dynamo model, in which the surface 
processes like the production of the 
poloidal field from the decay of active regions are important
as in the BL approach, but which has the same mathematical
sophistication as the PSKR approach. Such a dynamo model
should presumably account for both the equatorward migration
of sunspots and the poleward migration of the weak diffuse
field. An early step in this direction was taken by Wang, 
Sheeley, \& Nash (1991), who averaged over the radial direction to
obtain one-dimensional equations. More realistic two-dimensional
models have been developed by Choudhuri, Sch\"ussler, \& Dikpati (1995),
Durney (1995, 1996, 1997) and Dikpati \& Charbonneau (1999).

The mathematical theory of the PSKR approach is based on mean
field MHD, which leads to closed equations in the first order
smoothing approximation.  It is not clear if the implicit assumptions
in this mathematical theory are fully satisfied in any realistic
astrophysical situation.  However, if the assumptions are
satisfied, then the mathematical theory provides a completely
rigorous description of the dynamo process in the PSKR
approach.  To make a similar
rigorous formulation of the BL approach, we need to develop
a consistent mean field description of (i) the buoyant rise
of the toroidal flux to produce active regions and (ii) the
decay of the tilted active regions to produce the poloidal field.
In this paper, we focus our attention on comparing two possible 
formulations of the production of poloidal field from the decay of tilted active regions.  Since it is necessary to include magnetic buoyancy
to study this problem, we present some discussion of magnetic
buoyancy as well.

It was pointed out by Stix (1974) that the mathematical
formulation of the BL approach is in some ways analogous to
the $\alpha$-effect of the PSKR approach. 
Choudhuri, Sch\"ussler, \& Dikpati (1995) modeled
the decay of titled active regions to produce the poloidal field
by invoking an $\alpha$-coefficient which is concentrated near
the solar surface. Durney (1995, 1996, 1997) followed 
Leighton (1969) more closely and treated the same
by introducing a double ring of flux at the surface where
the eruption takes place.  Introducing an $\alpha$-coefficient
concentrated near the surface
is certainly a very approximate way of incorporating
the main idea of the BL approach into the mathematical theory
of the PSKR approach. Justifying this procedure rigorously is even
more difficult than justifying the $\alpha$-coefficient in the
PSKR approach.  However, this procedure produces the desired effect
of generating the poloidal field where we want to generate it.
If magnetic buoyancy is included in some way to bring the
strong toroidal field from the bottom to the top and then the
concentrated $\alpha$-effect acts on it, the net result is
similar to what happens in Durney's double ring method.
Since this procedure is easier to implement than Durney's
double ring method, one important question is whether this
procedure is at least as good as Durney's double ring method.
In this paper, we take a simple dynamo model and present
calculations done with both the methods.  We show that the
results are qualitatively similar.  It may be noted that it is
not our aim to build realistic models of the solar cycle in this
paper.  For example, we have presented a contrasting study
of these methods by assuming a differential rotation which
does not vary with latitude as in Choudhuri, Sch\"ussler, \& 
Dikpati (1995).  This simplification allows the specific features
of the two methods to be seen clearly.  A realistic differential
rotation makes the results immensely more complicated, which
we shall discuss in our next paper in which an attempt will
be made to model the solar cycle properly.

Durney (1995, 1995, 1997)
allowed flux eruption to take place only at one latitude at a
time.  In Durney's model, it is difficult to allow simultaneous
eruptions in a band of latitudes, which happens in the real
Sun.  The model of Choudhuri, Sch\"ussler, \& Dikpati (1995)
did not incorporate magnetic buoyancy and allowed the toroidal
field to be brought to the surface from the bottom by meridional
circulation.  To make comparisons with Durney's double ring
method, we now include magnetic buoyancy in that model by
allowing the magnetic field to erupt whenever it has a value
larger than a critical value.  It may be noted that incorporating
magnetic buoyancy in the PSKR approach was relatively easier,
since magnetic buoyancy there merely removed the flux from
the dynamo region and played the role of a dissipative process.
Some authors treated magnetic buoyancy by putting a simple loss
term in the dynamo equation (DeLuca \& Gilman 1986; Schmitt \&
Sch\"ussler 1989), whereas others included a general upward
flow due to magnetic buoyancy (Moss, Tuominen, \& Brandenburg
1990a, 1990b).  We have to go beyond such simple prescriptions
in a BL approach, where magnetic buoyancy is a more integral
part of the dynamo process and is not just a flux removal
mechanism.  In our BL model, magnetic buoyancy removes the flux
from the bottom layer where the toroidal field is generated
and then brings the flux to the top of the convection zone 
where the poloidal field is produced from it. Earlier, Choudhuri
\& Dikpati (1999) and Dikpati \& Charbonneau (1999) incorporated
the effect of magnetic buoyancy by including a dynamo source term
near the surface which is a product of the $\alpha$-coefficient
and the toroidal magnetic field {\it at the bottom\/} of the 
convection zone. 

From the observation that the following spots in active regions
appear at higher latitudes on the solar
surface, it is easy to figure out that $\alpha$ has to be positive 
in the northern hemisphere.  This is also clear from the expression
of $\alpha$-coefficient obtained by Stix [1974; Eq. (8)] by
recasting the equations of Leighton (1969). The positive sign of 
$\alpha$ gives
a new twist to the problem.  It is well known that the product
of $\alpha$ and the vertical gradient of differential rotation
has to be negative in the northern hemisphere for the equatorward
propagation of the dynamo wave (Parker 1955; see Choudhuri 1998,
\S16.6). Even if $\alpha$ and the velocity gradient are concentrated
in two different layers, this condition still remains valid
(Moffatt 1978, \S9.7). Since the vertical gradient of differential
rotation in the lower latitudes, as found by helioseismology,
is positive, its product with $\alpha$ is positive and one would
expect a poleward propagation of the dynamo wave.  It was
demonstrated by Choudhuri, Sch\"ussler, \& Dikpati (1995) that 
an equatorward propagation is still possible in this situation, if 
the time scale of meridional circulation is shorter than the time 
scale of diffusion between the layers of $\alpha$ and velocity shear.
This opens up the possibility of building models of the solar
dynamo in which we have a positive $\alpha$ near the surface
and a positive gradient of differential rotation at the bottom
of the convection zone.  The meridional circulation has to play
a very crucial role in such models in ensuring the desired behaviour.
While using Durney's double ring method, the signs of the magnetic
field in the two rings have to be chosen such that there is a
correspondence with the positive $\alpha$ situation.  With the
double ring method also, we found that the dynamo wave at the bottom
of the convection zone propagates equatorward only when there is
a strong meridional flow and propagates poleward when this flow
is switched off. 

In \S2 we discuss the details of our model. Then we go on to 
present our main results in \S3. Our conclusions are summarised in
\S4.

\section{The model}
We assume axisymmetry in all our calculations.  The magnetic
and velocity fields can be written as
\begin{equation}
\Bf = B \ep + \nabla \times (A \ep), 
\end{equation}
\begin{equation}
\vf = \vf_p + r \sin \theta \Omega \ep. 
\end{equation}
where $B$ and $A$ respectively represent the toroidal and poloidal
components of the magnetic field; $\Omega$ is the angular velocity,
and $\vf_p = v_r {\bf e}_r + v_{\theta} {\bf e}_{\theta}
$ is the meridional circulation.  We substitute
equations (1) and (2) in the induction equation
\begin{equation}
\frac{\pa \Bf}{\pa t} = \nabla \times (\vf \times \Bf) +
\eta \nabla^2 \Bf, 
\end{equation}
where $\eta$ is the coefficient of turbulent diffusion.  This 
gives 
\begin{equation}
\frac{\pa A}{\pa t} + \frac{1}{s}(\vf_p.\nabla)(s A)
= \eta \left( \nabla^2 - \frac{1}{s^2} \right) A + Q,
\end{equation} 
\begin{eqnarray}
\frac{\pa B}{\pa t} 
+ \frac{1}{r} \left[ \frac{\pa}{\pa r}
(r v_r B) + \frac{\pa}{\pa \theta}(v_{\theta} B) \right]
= \eta \left( \nabla^2 - \frac{1}{s^2} \right) B 
+ s(\Bf_p.\nabla)\Omega,
\end{eqnarray}
where $s = r \sin \theta$ and $\Bf_p = \nabla \times (A \ep)$.
We have added one extra term $Q$ on the right-hand side of equation (4),
which does not follow from the induction equation (3).  It is a
term which describes the generation of the poloidal field.  The
usual $\alpha \Omega$ dynamo is given by the equations (4) and (5),
where $Q$ is simply
\begin{equation}
Q = \alpha B. 
\end{equation}
To incorporate the effect of magnetic buoyancy and the decay of
tilted active regions, we have to  allow for changes in $B$ due 
to the rise of magnetic flux from the bottom of the convection
zone to the top and specify $Q$ appropriately.  Before describing
how we incorporate Durney's double ring method as well as our method of
concentrated $\alpha$-effect near the surface, let us discuss a few general
points which hold for both cases.

The equations (4) and (5) have to be solved in the northern
quadrant of the convection 
zone as usual (i.e.\ within $R_b =0.7 \Rs
\leq r \leq \Rs$, $0 \leq \theta \leq \pi/2$). The boundary
conditions are discussed in previous papers (Dikpati \& Choudhuri
1994; Choudhuri, Sch\"ussler, \& Dikpati 1995). They are
$$\mbox{At}\  \theta = 0: \ \ \ A = 0, \ B = 0,$$
 $$\mbox{At}\  \theta = \frac{\pi}{2}: \ \ \
\frac{\pa A}{\pa \theta} =0, \ B =0,$$
$$\mbox{At}\  r = R_b: \ \ \ A = 0,\ \frac{\pa}{\pa r}(rB) = 0, $$
$$\mbox{At}\  r = \Rs: \ \ \ B = 0,$$
the boundary condition for $A$ at the top $r = \Rs$ being that
it has to match a smooth potential field outside.  See \S3 of
Dikpati \& Choudhuri (1994) for a detailed discussion of how
this is implemented.

To solve equations (4) and (5) with these boundary conditions, we need to
specify $\eta$, $\Omega$, $\vf_p$ and $Q$.  As in Choudhuri, Sch\"ussler, \& Dikpati (1995), we assume the turbulent diffusion to have the constant
value $\eta = 1.1 \times 10^{7}$ m$^2$ s$^{-1}$.  For the angular
velocity $\Omega$ also, we use the same expression as used in that
paper:
\begin{equation}
\Omega = \Omega_0 \{0.9294 + 0.0353 \left[ 1 +
\er \left( \frac{r - r_3}{d_3} \right) \right] \} 
\end{equation}
with $r_3 = 0.7 \Rs$, $d_3 = 0.1 \Rs$, $\Omega_0 = 2.7 \times
10^{-6}$ s$^{-1}$.  This latitude-independent angular velocity
roughly corresponds to the helioseismologically determined 
rotation profile near the solar equator, with $\pa \Omega/
\pa r$ positive. For the meridional circulation $\vf_p$, we
again use the expression used previously (Dikpati \& Choudhuri
1995; Choudhuri, Sch\"ussler, \& Dikpati 1995).  In other 
words, we take
\begin{equation}
\rho \vf_p = \nabla \times (\psi \ep)
\end{equation}
with $\psi$ given by
\begin{eqnarray}
\psi r \sin \theta = \psi_0 \sin \left[ \frac{\pi (r - R_b)}
{(\Rs - R_b)} \right] \{ 1 - e^{- \beta_1 r \theta^{\epsilon}}
\}\times\{1 - e^{\beta_2 r (\theta - \pi/2)} \} e^{
-((r -r_0)/\Gamma)^2}
\end{eqnarray}
and $\rho$ given by
\begin{equation}
\rho(r) = C \left( \frac{\Rs}{r} - \gamma \right)^m 
\end{equation} 
The values of the parameters used are $\beta_1 = 1.4 \times
10^{-8}$ m$^{-1}$, $\beta_2 = 2.7 \times 10^{-8}$ m$^{-1}$,
$\epsilon = 2.0000001$, $r_0 = (\Rs - R_b)/5$, $\Gamma =
3.3 \times 10^{8}$ m, $\gamma = 0.9$, $m = 3/2$.
The pattern of meridional circulation for these values
of parameters is shown in Figure~3a of Dikpati \& Choudhuri
(1995).  The amplitude of the meridional circulation is fixed
by taking $\psi_0 / C = - 7.9 \times 10^{8}$ m$^2$ s$^{-1}$, 
which corresponds to a maximum
surface velocity ($v_0$) of about $7.0$ m s$^{-1}$ 
in the mid-latitudes. 

In the paper
of Choudhuri, Sch\"ussler, \& Dikpati (1995), the source 
term $Q$ was given by
equation (6) with $\alpha$ taken in the form
\begin{eqnarray} 
\alpha =\frac{\alpha_0}{1 + B^2} \cos \theta \frac{1}{4}
\left[ 1 + \er \left(\frac{r - r_1}{d_1} \right) \right]
\times
\left[ 1 - \er \left(\frac{r - r_2}{d_2} \right) \right].
\end{eqnarray}
The parameters are $r_1 =0.95 \Rs$, $r_2 = \Rs$, $d_1 = d_2 =
0.025 \Rs$, making sure that the $\alpha$ effect is concentrated
in the top layer $0.95 \Rs \leq r \leq \Rs$.  The $\alpha$-quenching
factor $1 + B^2$ included in the denominator helps the system to
relax to periodic solutions with amplitude $B \sim 1$.  
This essentially means that we are choosing the unit of $B$
in such a way that the nonlinear feedback becomes important
when $B$ is of order unity or larger.
Choudhuri, Sch\"ussler, \& Dikpati (1995) took 
$\alpha_0 = 3$ m s$^{-1}$ and found that it
gave rise to marginally critical oscillations.  When magnetic buoyancy
is included, we find that this value of $\alpha_0$ often gives
decaying solutions.  To ensure that the solutions do not decay,
we take $\alpha_0 = 10$ m s$^{-1}$ in most of the calculations in
the present paper. 

\subsection{Incorporating the double ring}

After time intervals $\tau$, we find the co-latitude $\ter$ where the
toroidal field is maximum and allow the flux to erupt above in the
form of the double ring, if this maximum value exceeds a specified 
critical field $\Bc$.
Following Figure~1 of Durney (1997), we show the two emergent flux rings
in Figure~1. One ring of positive magnetic field $K/ \sin \theta$
is put between the co-latitudes $\theta_1$, $\theta_2$, whereas
the other ring of negative magnetic field $-K/ \sin \theta$ is
between $\theta_3$, $\theta_4$.  The factor $\si$ ensures that
the flux through one ring balances the flux through the other
ring.  In Durney's notation, $\theta_1$, $\theta_2$, $\theta_3$
and $\theta_4$ will be
$$\theta_1 = \ter - \frac{\chi + \Lambda}{2},$$
$$\theta_2 = \ter - \frac{\chi - \Lambda}{2},$$
$$\theta_3 = \ter + \frac{\chi - \Lambda}{2},$$
$$\theta_4 = \ter + \frac{\chi + \Lambda}{2}.$$
As in Durney (1997), we make the somewhat unphysical assumption
that these rings extend only from $R_{\odot}$ to $R_{\odot} -
\Delta r$, where the field lines end abruptly. At the time of
eruption, then, in the region $\Ro - \Delta r \leq r \leq \Ro$,
we put the magnetic field

\begin{eqnarray}
\Delta B_r  & = & \frac{K}{\si}\; \; \mbox{if} \; \; \theta_1 \leq \theta 
\leq \theta_2 \nonumber \\ 
          & = & \frac{- K}{\si}\; \; \mbox{if} \; \; \theta_3 \leq  \theta
\leq \theta_4 \nonumber \\
          & = & 0 \; \; \mbox{elsewhere} 
\end{eqnarray}
Putting this magnetic field is equivalent to adding the vector
potential $\Delta A $ given by
$$\Delta B_r = \frac{1}{r \si} \frac{\pa}{\pa \theta} (\si \Delta A),$$
from which
\begin{equation}
\Delta A = \frac{r}{\si} \int_0^{\theta} \sin \theta' \Delta B_r
d \theta' 
\end{equation}
if we do not consider the variation of $\Delta A$ in $r$. Substituting
for $\Delta B_r$ from (12), we conclude that $\Delta A$ can be
non-zero only in the range $\Ro - \Delta r \leq r \leq \Ro$,
where we have
\begin{eqnarray}
\Delta A \si & =  0 \; \; & \mbox{for} \;\; 0 \leq \theta \leq \theta_1 
\nonumber \\
& =   \Ro K (\theta - \theta_1) \; \; & \mbox{for} \;\; \theta_1 \leq \theta \leq \theta_2 \nonumber \\
& =  \Ro K (\theta_2 - \theta_1) \; \; & \mbox{for} \;\; \theta_2 \leq \theta \leq \theta_3 \nonumber \\
& =  \Ro K [(\theta_2 - \theta_1) - (\theta - \theta_3)]
\; \; & \mbox{for} \;\; \theta_3 \leq \theta \leq \theta_4
\nonumber \\
& =  0 \; \; & \mbox{for} \;\; \theta_4 \leq \theta \leq \frac{\pi}{2} 
\end{eqnarray}
Adding this $\Delta A$ to $A$ leads to a discontinuity in $A$
at $\Ro - \Delta r$.  Durney (1997) writes, ``Such an expression
for the vector potential generates latitudinal magnetic fields
(associated with the closure of magnetic lines of force)'', but
also claims that these discontinuities ``are numerically
inconsequential''. 

Durney (1995, 1996, 1997) took the separation between the rings,
$\chi$, to be proportional to $\cos \ter$.  To keep the numerical
computations simpler, we instead take $K$ appearing in (14) 
to be proportional to
$\cos \ter$, while keeping the separation between the rings fixed.
This has the same physical effect, except at the low latitudes
where the two rings may overlap.  We, however, find that flux
eruption remains restricted to higher latitudes where this overlap
is unimportant. If we take $K$ to be proportional to the toroidal magnetic
field $B$ at the bottom (from which the flux rings originate), 
then our problem becomes linear in magnetic field and one
has to make many runs to find the marginally growing solution.
We circumvent this problem by including something like $\alpha$-quenching
in the following fashion:
\begin{equation}
K = K' \frac{B_{\rm max} \cos \ter}{1 + |B_{\rm max}|^2 },
\end{equation}
where $B_{\rm max}$ is the toroidal magnetic field at the bottom
of the convection zone at the latitude where it is maximum.
The justification behind this is the fact that a stronger toroidal field
is less affected by the Coriolis force (D'Silva and Choudhuri 1993;
Howard 1993) and hence is less efficient in generating the poloidal
field.  It can easily be seen from (14) and (15) that $K'$ is
a dimensionless quantity.

It is seen on the solar surface that the magnetic field of the
higher-latitude sunspot is positive when the toroidal field
underneath the surface is positive.  It should be clear from
(12) and (15) that this is achieved by taking $K'$ positive,
which is the case in all our calculations.  It now follows
from (14) and (15) that a positive $B$ at the bottom of the
convection zone would imply a positive increment in $A$ at
the surface where the magnetic flux emerges.  This is something
like a positive $\alpha$-effect, which obviously corresponds
to a positive value of $K'$.

We solve (4) and (5) with differential rotation
and meridional circulation as given by (7)--(10).  The source
term $Q$ in (4) is given by (6) and (11).  However,
in addition to this usual source term, we 
allow for possible changes in the value of
$A$ abruptly, in the double ring regions of the surface, at intervals
of $\tau$, to take account of magnetic buoyancy.
We run our code to find the maximum value of $B$ after 
intervals $\tau$.  If this exceeds the critical field $\Bc$ and 
occurs at the co-latitude $\ter$, then
we consider two rings situated on two sides of this co-latitude
and add $\Delta A$ as given by (3) to $A$.  The control parameter
in our problem is $K'$ appearing in (4).  When $K'$ is zero,
there is no double ring formation and we get the model of
Choudhuri, Sch\"ussler, \& Dikpati (1995).  
On the other hand, when $K'$ is sufficiently large,
the net effect of double ring formation at intervals of $\tau$
becomes much more important than the source term $Q$ in (4)
and we have the model of Durney (1997).  Thus, in the two
opposite limits of the control parameter $K'$, our model is
respectively reduced to the models of Choudhuri, Sch\"ussler, \& 
Dikpati (1995) or Durney (1997).

\subsection{Concentrated $\alpha$-effect with magnetic buoyancy}

We wish to argue that the double ring method is similar to allowing
magnetic flux to rise due to magnetic buoyancy and then letting
the $\alpha$-effect concentrated near the surface to act on it.
In this method also, we solve (4) and (5) in conjunction with
(6)--(11).  However, instead of having double ring formations at
intervals of $\tau$ (leading to abrupt changes of $A$ as seen from
[14]), we now allow $B$ to change abruptly at intervals of $\tau$
to take account of flux rise due to magnetic buoyancy.  This is
done in the following way.

We assume that the toroidal field $B$ becomes
buoyant when its value crosses a critical value $\Bc$. After
intervals of time $\tau$, we check if $B$ has become larger than
$\Bc$ at certain points.  Then, at those points, $B$ is reduced
by a factor $1-f$, i.e. $$B \rightarrow B (1 - f).$$
The flux removed from these points is taken vertically above
and deposited near the surface by increasing $B$ there in such
a fashion that the total flux remains conserved in the transfer
process.  Since the equations are numerically solved on a $N \times
M$ grid, the simplest procedure is to deposit all the flux at the
grid point just below the surface.  For example, if $B$ crosses
$\Bc$ only at one grid point on the radial line at a fixed latitude,
then we have to decrease $B$ by $f B$ there and the toroidal
field at the grid point just below the surface has to be increased
by an amount $f' B$.  Since the grid size at the surface corresponds
to a greater distance in the latitudinal direction than that at 
the bottom, we need to take $f' = f (R_i/R_f)$ to ensure the
conservation of magnetic flux (here, $R_i$ is the radius near the bottom
where the flux is depleted and $R_f$ is the radius near the
surface where the flux is deposited). We have also made some runs in which the
flux taken up from one grid point is distributed within a 
few grid points near the top instead
of all the flux being deposited in one grid point, and the results 
turn out to be qualitatively similar.  The strength
of magnetic buoyancy is increased by increasing the control
parameter $f$. In the limit $f=0$, we get back the model of 
Choudhuri, Sch\"ussler, \& Dikpati (1995), in which there was no
magnetic buoyancy and the toroidal field was brought to the surface
by the meridional circulation.  When $f$ is made sufficiently large
(even though it has to remain less than 1), magnetic buoyancy is
found to dominate and the system has a limiting behavior.

Compared to the double ring method, this method has some attractive
features.  Firstly, here the eruption at any instant takes place
over a range of latitude rather than at one point as in the double
ring method.  This corresponds to the real Sun more closely.  It
is not easy to extend the double ring method to handle simultaneous
flux eruptions at more than one point.
If we simultaneously put several double rings in a range of latitudes,
then the positive ring of an intermediate double ring will cancel
with the negative ring of the next double ring and we shall be left
with a positive ring and a negative ring at a wide separation.  It
follows from (14) that this will mean adding to $A$ over a wide range
of latitude.  This would make the model more similar to the mean field
model and the special character of the original double ring model
would be completely lost.  
Also, we now allow for
the toroidal flux to be depleted at the bottom of the convection zone
due to magnetic buoyancy.  As we shall argue later, we believe this
to be quite important.  In fact, we shall present some results with the double ring method with the toroidal flux at the bottom depleted parametrically.
 
\section{Results}

We now present and compare results obtained by the two methods described
above.  As we saw, $K'$ and $f$ happen to be the respective control
parameters in these two methods.  On setting these control parameters
equal to 0, both these methods are reduced to the model of Choudhuri, 
Sch\"ussler, \& Dikpati (1995, hereafter CSD model). All 
our calculations are done on a $64 \times 64$ grid. We allow the 
eruptions to take place after times $\tau = 8.8 \times 10^5$ s and 
use a value $\Bc = 1$ for the critical field in all our calculations. When we start our calculations with any arbitrary magnetic field configuration, 
the code relaxes to
a periodic solution for a proper set of parameters.  What we discuss 
below are properties of such relaxed periodic solutions.

\subsection{Results with the double ring method}

Durney (1997) did not allow the toroidal flux to be depleted at the
bottom of the convection zone due to magnetic buoyancy.  To study the
effect of flux depletion, we present some calculations in which we
allow flux depletion in the following simple manner.  At the times
of eruption after interval $\tau$, we find out at which point the
toroidal field has the maximum value $\Bm$ ($>\Bc$).  While putting 
the two flux rings at the top, we also decrease $\Bm$ by an amount $f_d \Bm$ at the maximum point.  Then $f_d$ becomes a second parameter in the
problem in addition to $K'$ in our problem.   After finding the
co-latitude $\ter$ where the toroidal field is maximum, the next
two poleward grid points are taken as $\theta_1$, $\theta_2$, and
the next two equatorward grid points are taken as $\theta_3$,
$\theta_4$. The flux rings are assumed to go 3 grid points
deep (i.e.\ $\Delta r$ is taken 3 grid points below the surface).

Figure 2 shows how the dynamo period $T_d$ changes with the parameter
$K'$ when $f_d$ is held constant.  The different curves correspond
to different values of $f_d$.  When we go to the limit of CSD model
by putting $K'=0$, we find the period to be 66 yrs.  When $f_d = 0$
(i.e.\ there is no flux depletion at the bottom), we find that the
change in the period with $K'$ does not follow any particular trend.  
$T_d$ at first increases slightly with increasing $K'$ and then comes down to a value close to that of the CSD model.  This behaviour for $f_d =0$
may result from the fact that in this case we are actually creating flux (in the form of erupted double rings) without any depletion. More meaningful behaviour follows for the
other values of $f_d$ (such as 0.25, 0.5, 0.75). The period decreases
with increasing $K'$ and tends to saturate at some asymptotic value
for large $K'$.  To understand what is happening, let us look at
Figure~3 which shows the evolution of magnetic field during a 
half-period for the case $K' = 1000$, $f_d = 0.5$.  In the plots of poloidal field, we have indicated the latitudes of last flux eruption
with small arrows.  However, the individual
double rings are not usually discernable.  That is
not surprising.  Flux eruption in the form of double rings
keeps occuring at intervals of $\tau$.  Hence the latest double
ring is merely superposed on the field created by the previous
double rings and does not stand out against the background of
previously created field. On looking at the plots of the toroidal
field, it is clear that the toroidal field keeps weakening as
we go to lower latitudes.  This weakening of toroidal field at
lower latitudes becomes more prominent as we make $f_d$ larger.
This implies that flux eruption never takes place at very low
latitudes and the dynamo process is basically confined to higher
latitudes. Since it takes less time to transport magnetic flux
through a limited range of latitudes, the dynamo period is shorter
for non-zero $f_d$. In combination with this effect, an increasing 
$K'$ will make the erupted double rings stronger, thus recycling
toroidal flux to poloidal flux more efficiently. This reduces the
time period of the dynamo as compared to the period in the limit 
of the CSD model, in which the toroidal field is brought to the 
surface by the meridional flow only near the equator and the whole 
range of latitudes is involved. It may be noted that Durney (1997) 
did not present any plots of magnetic field configurations in his paper. 
However, we do get a deeper insight into the problem by looking at such field configuration plots. For example, note that the direction of the
poloidal field (clockwise or anti-clockwise) starts reversing at the 
time when we have an extended belt of strong toroidal field.

Durney (1997) has presented several plots showing how the 
eruption latitude changes with time (Figures~7--10 in his
paper).  We present a similar plot in Figure~4 for the case 
$K' = 1000$, $f_d = 0$, corresponding to no flux depletion at
the bottom as in the calculations of Durney (1997). Here, we see 
that eruptions continue near the pole for some time at the beginning 
of a cycle and then progressively move to lower latitudes.
This plot looks very much like the plots presented 
by Durney (1997) --- especially his Figure~7.  This is
certainly very reassuring, since the numerical techniques
employed by us and by Durney (1997) are completely different.
Apart from the production of the double rings, 
our code allows for the toroidal flux
to be brought to the surface by meridional circulation and
then to be acted upon by $\alpha$-coefficient (an effect not
present in Durney's calculations).  However,
when $K'$ is made as large as 1000, this effect is insignificant.
In fact, we made some runs with $\alpha = 0$ and found that
the results for zero or non-zero $\alpha$ are virtually
indistinguishable when $K' = 1000$. For example, the plots of 
eruption latitude against time and the butterfly diagrams look 
identical in both the cases.

We have already mentioned that a positive $K'$ is like a
positive $\alpha$-effect concentrated near the surface. 
Choudhuri, Sch\"ussler, \& Dikpati (1995) 
showed that a positive $\alpha$ concentrated
near the surface leads to a poleward propagation of the dynamo
wave when the meridional flow is switched off.  We find 
exactly the same result in the double ring approach with
positive $K'$ if we switch off the meridional flow.  
Figure~5 shows a time-latitude plot of the toroidal field
at the bottom of the convection zone with meridional flow
for the case $K'=1000$, $f_d = 0.5$, whereas Figure~6 is a similar
plot without meridional flow keeping all the other parameters
the same.  We see clear indication of poleward migration in 
Figure~6.

\subsection{Results for concentrated $\alpha$ with buoyancy}

For contrast, we now present results obtained by the method
described in \S~2.2.  As we have seen, the control parameter
in this problem is $f (<1)$, which measures the strength of magnetic
buoyancy.  Figure~7 shows how the dynamo period changes on increasing
$f$.  As in Figure~2, we begin with a period of 66 yrs in the limit
$f= 0$ corresponding to the CSD model.  On making the effect of 
buoyancy stronger (by increasing $f$), the flux transport (from the 
bottom of the convection zone to the top) takes place more 
efficiently and also the toroidal flux gets depleted quickly. This 
results in the dynamo period reducing with increasing $f$, until it 
reaches an asymptotic value of about 25 yrs.  We may point out here 
that we did some runs for this method without depleting the field at 
the bottom, which would correspond to the case $f_d =0$ for the double 
ring method. We found that even in this case, there is no decrease in 
the time period with increasing $f$ (which in this case corresponds 
to only field addition at the top) and $T_d$ more or less hovers 
around the CSD limit of 66 years.
 
Since the two methods discussed by us are sufficiently different,
it is not obvious which value of $K'$ in the first method would
correspond to a certain value of $f$ in the second method.
In both methods, however, the dynamo periods saturate to
asymptotic values when these control parameters are sufficiently
large.  So the most sensible thing is to compare results of
the two methods when the control parameters are large enough
to ensure that the dynamo period has the asymptotic value.
Figure~8 shows the time evolution of the magnetic field
during a half period for the parameters $f = 0.05$ (i.e.\
the magnetic buoyancy is strong enough to saturate the period to
its asymptotic value).  On comparing with Figure~3, we find that 
the broad features of the magnetic field distribution are very
similar.  The main difference is that one sees some toroidal field
distributed near the top of the convection zone in Figure~8, whereas
such fields are not present in Figure~3. The reason behind this is
obvious.  In the double ring method, we directly put double rings
above regions of strong toroidal field and this contributes directly
to the poloidal field.  When we introduce the intermediate step of
the toroidal field first rising due to buoyancy and then being acted
upon by $\alpha$-effect, then we get toroidal field at the top of
the convection zone also, as in Figure~8.  The other difference
between Figures~3 and 8 is that often the field lines in Figure~3 in some
places (especially near the surface) are not as smooth as the field
lines are everywhere in Figure~8.  This is certainly due to double
ring formations in Figure~3, which are concentrated local effects.
As in Figure~3, here also we find that the direction of poloidal field reverses at around the time the strong toroidal field 
belt is maximally extended. 

Finally Figure~9 presents a time-latitude plot of the toroidal field
at the bottom for the same case which is presented in Figure~8.
Again, this figure looks qualitatively similar to Figure~5, the main 
difference being the fact that the toroidal field has become much weaker
near the equator in Figure~8 due to more efficient flux depletion at the bottom, 
which takes place naturally in this method.

\section{Conclusion}
 
Following Choudhuri, Sch\"ussler, and Dikpati (1995) and Durney (1997),
we build a hybrid model of the solar dynamo, in which the 
best features of both the PSKR and the BL approaches are combined.  The
aim is to include the surface processes emphasized in BL models into
a model as suitable for detailed quantitative study as the PSKR
models.  We study two possible methods of achieving this.  One is to
introduce double rings above the region where the toroidal field is
maximum, as done by Durney (1995, 1996, 1997).  The second method is
to make the toroidal field rise when it is above a critical value and
then allow it to be acted upon by an $\alpha$-coefficient concentrated
near the surface. It is reassuring that the results obtained by the two 
methods are qualitatively similar.  
 
We believe that the depletion of toroidal flux by magnetic
buoyancy is an important process. Flux
tube calculations (Choudhuri \& Gilman 1987; Choudhuri 1989;
D'Silva \& Choudhuri 1993; Fan, Fisher, \& DeLuca 1993; 
Caligari, Moreno-Insertis, \& Sch\"ussler 1995)
suggest that the toroidal field at the bottom of the convection
zone has a value of $10^5$ G---much stronger than the equipartition
value.  After the belt of strong toroidal field reaches the
equator, it disappears and the next half-cycle of the dynamo
begins.  If the field is so strong, then turbulent diffusion
will be completely suppressed and will not be effective in
destroying the strong toroidal field.  The only way to annihilate
this belt of strong toroidal field is to expect magnetic buoyancy
to deplete its strength sufficiently by the time this belt
propagates to the equator. In our second method, this flux
depletion automatically takes place.  In the double ring method,
we have included the possibility of toroidal flux depletion as
an extra effect, which was not taken into account by Durney (1997).
When the toroidal flux is depleted appropriately, both the methods
make the period of the dynamo decrease on increasing the control
parameters ($K'$ or $f$) and saturate at some asymptotic value.
This decrease of period is due to the efficient and rapid
transport of toroidal flux by magnetic buoyancy.  It is true
that the decrease of period is more pronounced in the second
method, where the flux depletion is more prominent.  However,
a difference by a factor 2 or 3 in the asymptotic period is
probably not such a significant uncertainty compared to many
other factors.  The magnetic field configurations obtained in
the two methods, as seen in Figures~3 and 8, are also quite
similar, with the poloidal field reversing its direction for 
the same configuration of the toroidal field.  Then, very 
importantly, the dynamo wave is found to propagate poleward 
when the meridional circulation is switched off in the double 
ring method.  
In other words, the double ring method with positive $K'$ has 
characteristics quite similar to a model with positive 
$\alpha$-effect concentrated near the surface. The results 
obtained by the two methods 
are not exactly identical.  However, given the many uncertainties 
plaguing the solar dynamo theory at the present time, representing 
the generation of poloidal field near the surface by a concentrated 
$\alpha$-effect acting on erupted toroidal field seems like a good 
enough approximation. 

We should point out that there are several logistic problems
in numerically handling the double ring, which are not there
if we use an $\alpha$-effect instead.  Firstly, 
to properly create rings of latitudinal size similar to sunspot
size with appropriate separation, one has to either use at
least 500 grid points in the $\theta$ direction or use a special
code which employs a finer mesh in the region where eruption 
takes place. Durney (1997) used $101 \times 101$ grid, which corresponds
to a grid size of about 11,000 km in the latitudinal direction at the Sun's
surface.  The width of the double ring has to be at least 4
times this, i.e.\ about 44,000 km---definitely inadequate to
resolve the north-south polarity separation of a typical active
region.  To
ensure whether our results have converged with respect to grid
size, we repeated some calculations on $32 \times 32$ grid and
compared the results with those obtained on our usual
$64 \times 64$ grid.  We found that results obtained by our
second method of concentrated $\alpha$-effect were so close 
in the two cases that various plots
looked indistinguishable.
However, results obtained by the double ring method, in which
important source terms are taken at the limit of grid resolution,
while remaining qualitatively similar on halving the grid size,
showed some changes.  Our grid size is comparable to what other
researchers (Durney 1997, Dikpati \& Charbonneau 1999) have used
on similar problems.  We believe that the grid size has to be
reduced considerably to properly resolve double rings and
to give results completely invariant with grid size.  
Since the ring
separation was at the limit of grid resolution, we kept the
ring separation fixed and made our constant $K$ proportional
to $\cos \theta_{\rm er}$ (see [15] and the discussion preceding
that).  Durney (1997) claims to have made the ring separation
proportional to $\cos \theta_{\rm er}$, but never explains in his paper
how this could be done with only 101 grid points in the $\theta$
direction.  Another important
consideration is that the double ring method is easy
to implement when we allow flux eruption only at one point at
one time, but it is not easy to generalize if multiple flux
eruptions are allowed.  In reality, we find that, at a certain
time, several active regions emerge in a belt of latitudes---with
the different active regions usually separated in longitude.
If one could use an appropriately resolved 3D code in which 
active regions of realistic
size were made to emerge in different latitudes and longitudes,
then certainly that would have been a much more satisfactory
calculation than what we are doing now.  We hope that future
computers will be used by researchers more numerically capable
than us to tackle this problem. When one uses rings
to replace active regions through an averaging over longitude
and uses a grid not fine enough to resolve individual sunspots,
one already introduces some drastic averaging.  Introducing
an $\alpha$-coefficient concentrated near the surface instead
of using double rings may not be such a big step after that.

Let us end with a comment on what we mean by a Babcock-Leighton
model, since this term often creates some confusion.  Babcock (1961)
and Leighton (1969) emphasized the surface process of poloidal
field generation from tilted active regions---in contrast
to the usual mean field MHD where the poloidal field is supposed
to be produced in the interior region of turbulence (see, for
example, Choudhuri 1998).  Hence any dynamo model in which
the poloidal field is generated in a thin layer near the solar
surface should be called a Babcock-Leighton model.  Dikpati \&
Charbonneau (1999) also use the term in this sense.  Durney (1995, 1996,
1997) followed Leighton (1969) more closely in incorporating
the Babcock-Leighton idea through the double ring method.
Introducing a phenomenological $\alpha$-coefficient concentrated
near the surface is another way of representing the Babcock-Leighton
idea.  One should, however, be careful not to interpret this
$\alpha$-coefficient in the way it is interpreted in the mean
field MHD.  For example, the $\alpha$-coefficient here is not
obviously related to the average helicity of turbulence as in
mean field MHD (see, for example, Choudhuri 1998, \S16.5). This 
coefficient merely provides a phenomenological
description of the production of poloidal field from the decay
of tilted active regions, which is obvious in the formulation
of the BL model by Stix (1974).  Wang \& Sheeley (1991) also referred 
to this process as an ``$\alpha$-effect'' in exactly the same
sense as us, even though they never
used the symbol $\alpha$ in their actual equations!

\acknowledgements{
We would like to thank Paul Charbonneau, Bernard Durney, 
Gene Parker and an anonymous referee for valuable suggestions.} 

\pagebreak

\pagebreak

\begin{figure}[h]
\epsfig{file=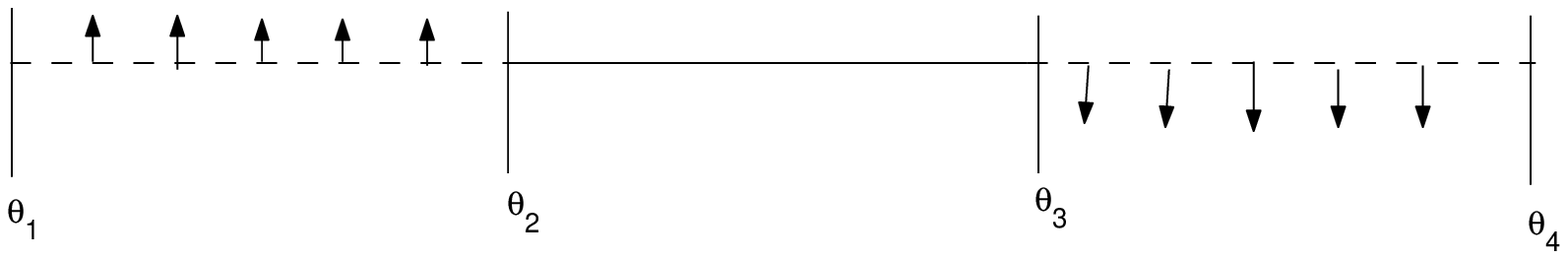,width=8.0cm,angle=0}
\caption{A sketch of the erupted double ring at the solar surface.}
\end{figure}     

\begin{figure}[h]
\epsfig{file=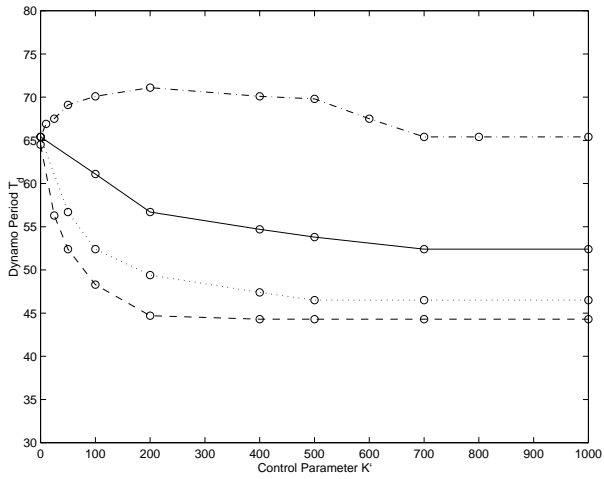,width=8.0cm,angle=0}
\caption{The variation in the dynamo period (in units of years)
with the control parameter $K'$ for four different $f_d$ values. The
dash-dotted line corresponds to $f_d = 0$, the solid line to 
$f_d = 0.25$, the dotted line to $f_d = 0.5$ and the dashed line
to $f_d = 0.75$.}   
\end{figure} 

\begin{figure}[h]
\epsfig{file=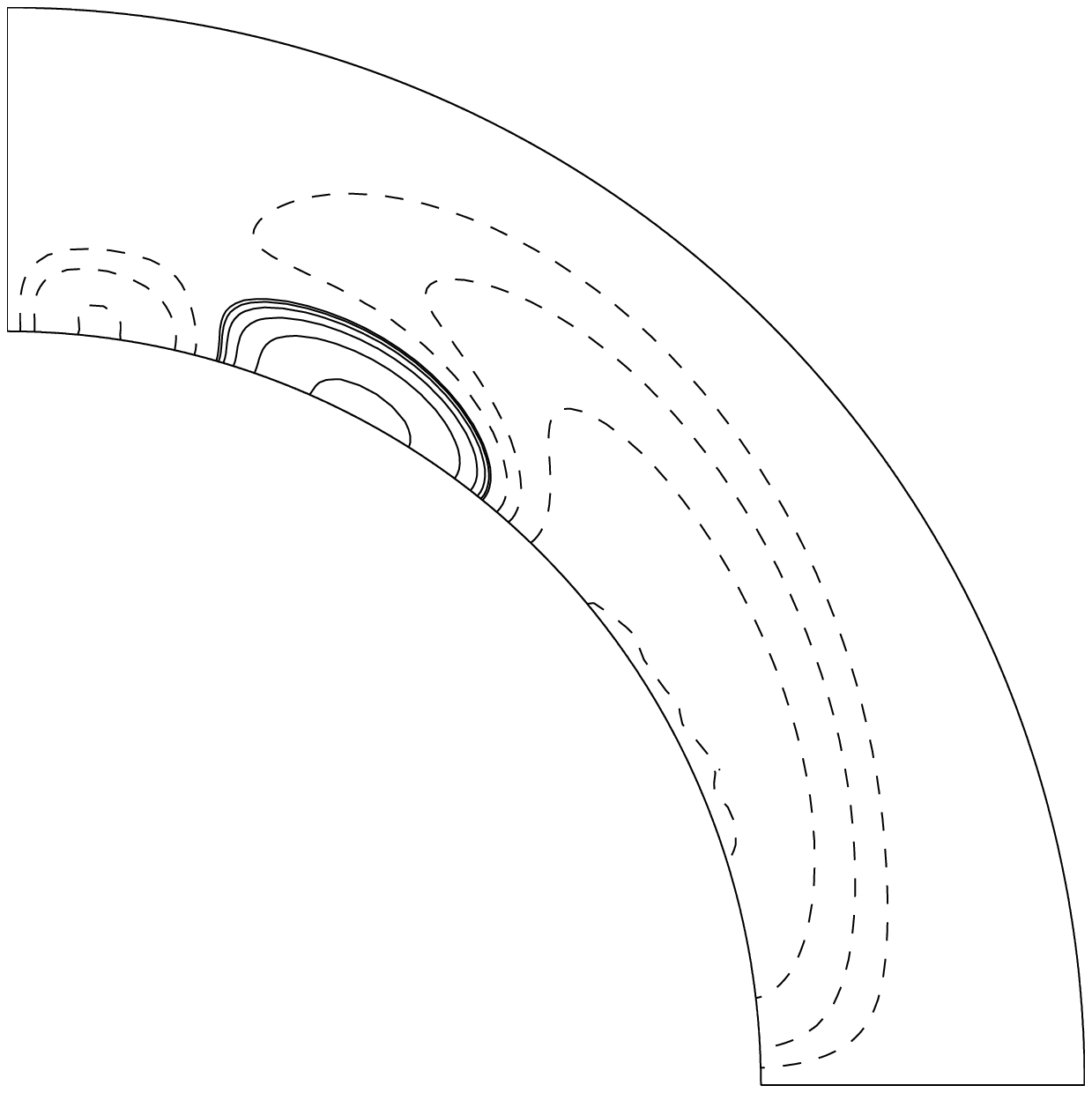,width=4cm,angle=0}
\epsfig{file=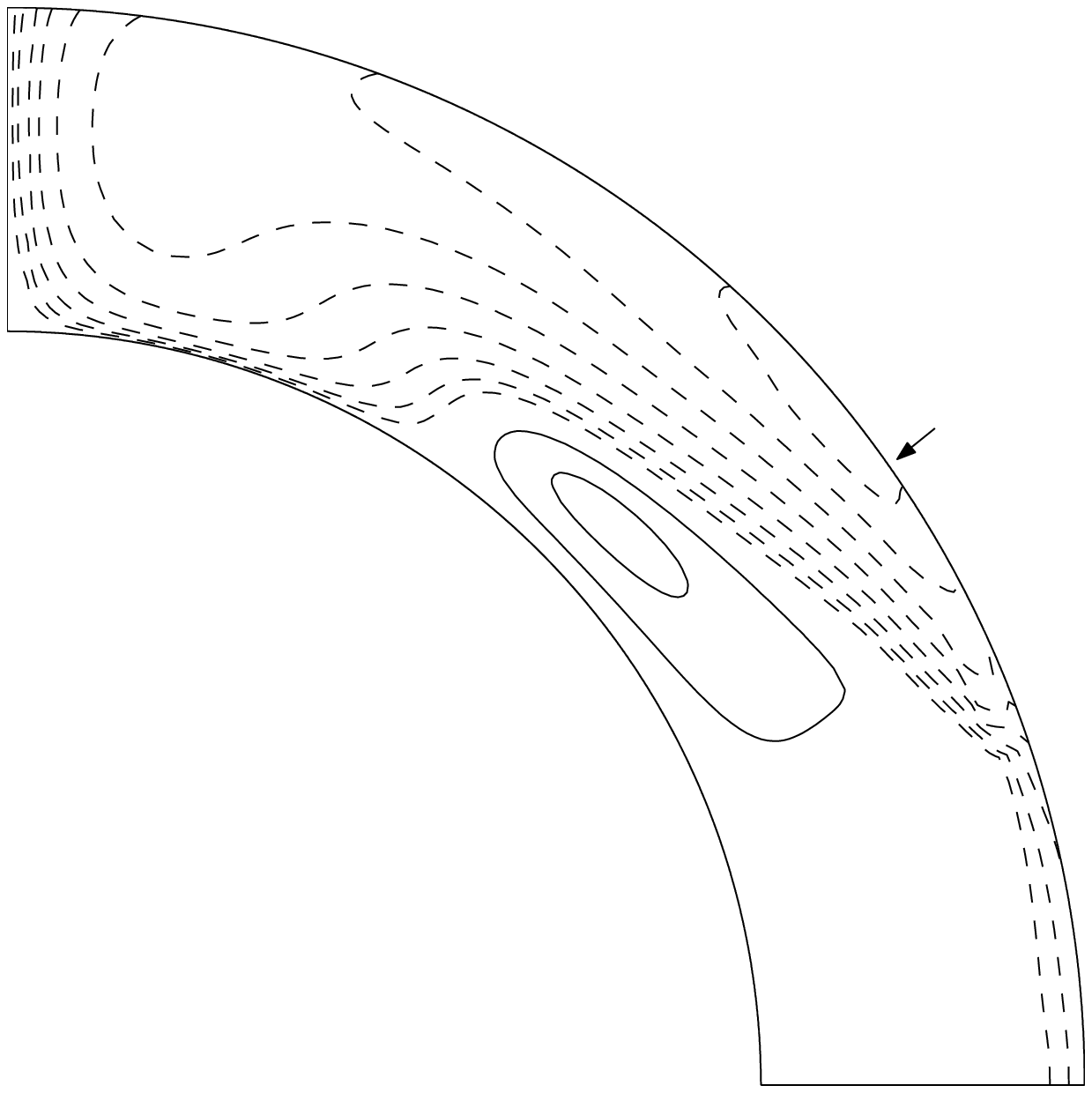,width=4cm,angle=0}
\vfill
\epsfig{file=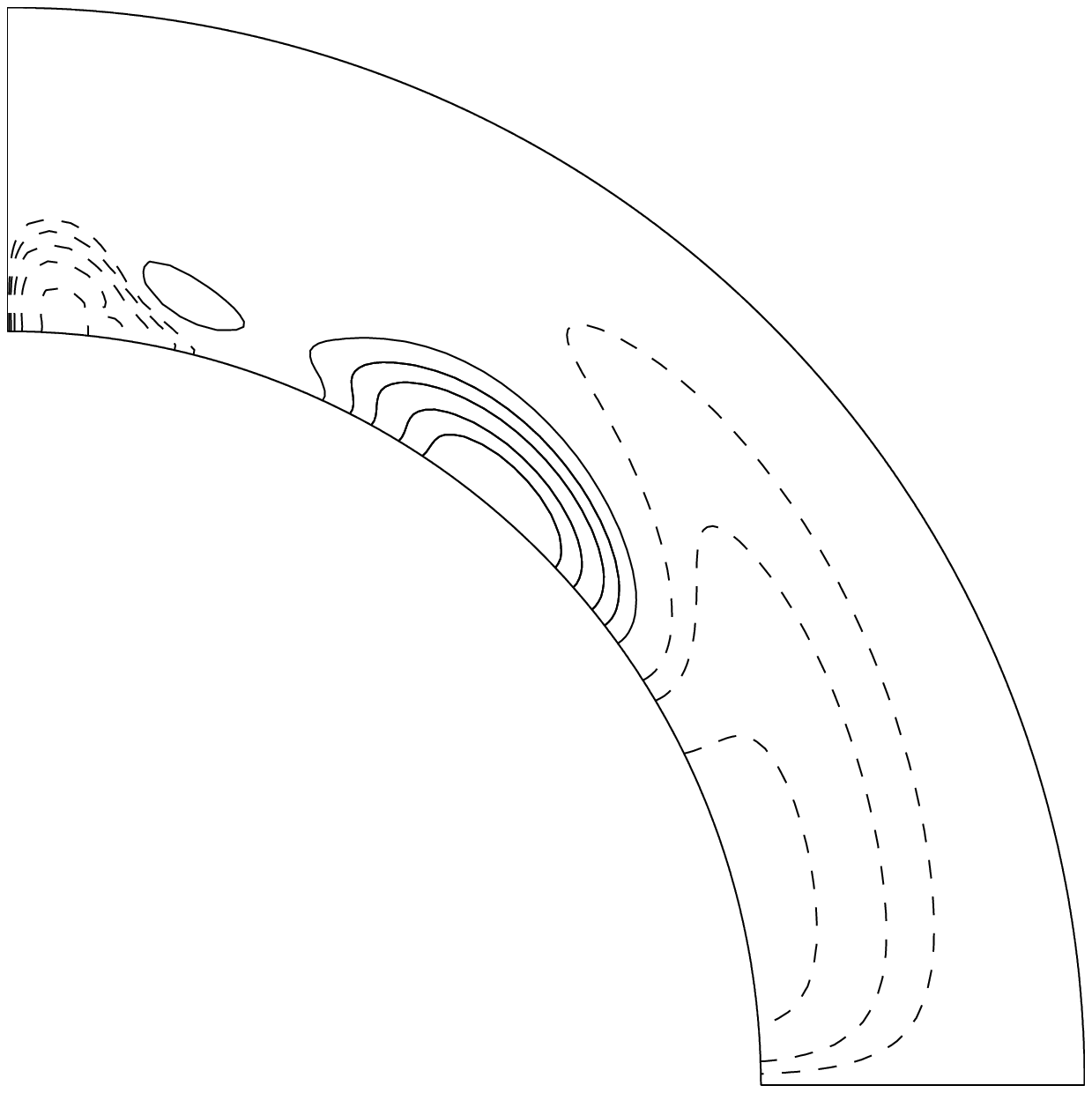,width=4cm,angle=0}
\epsfig{file=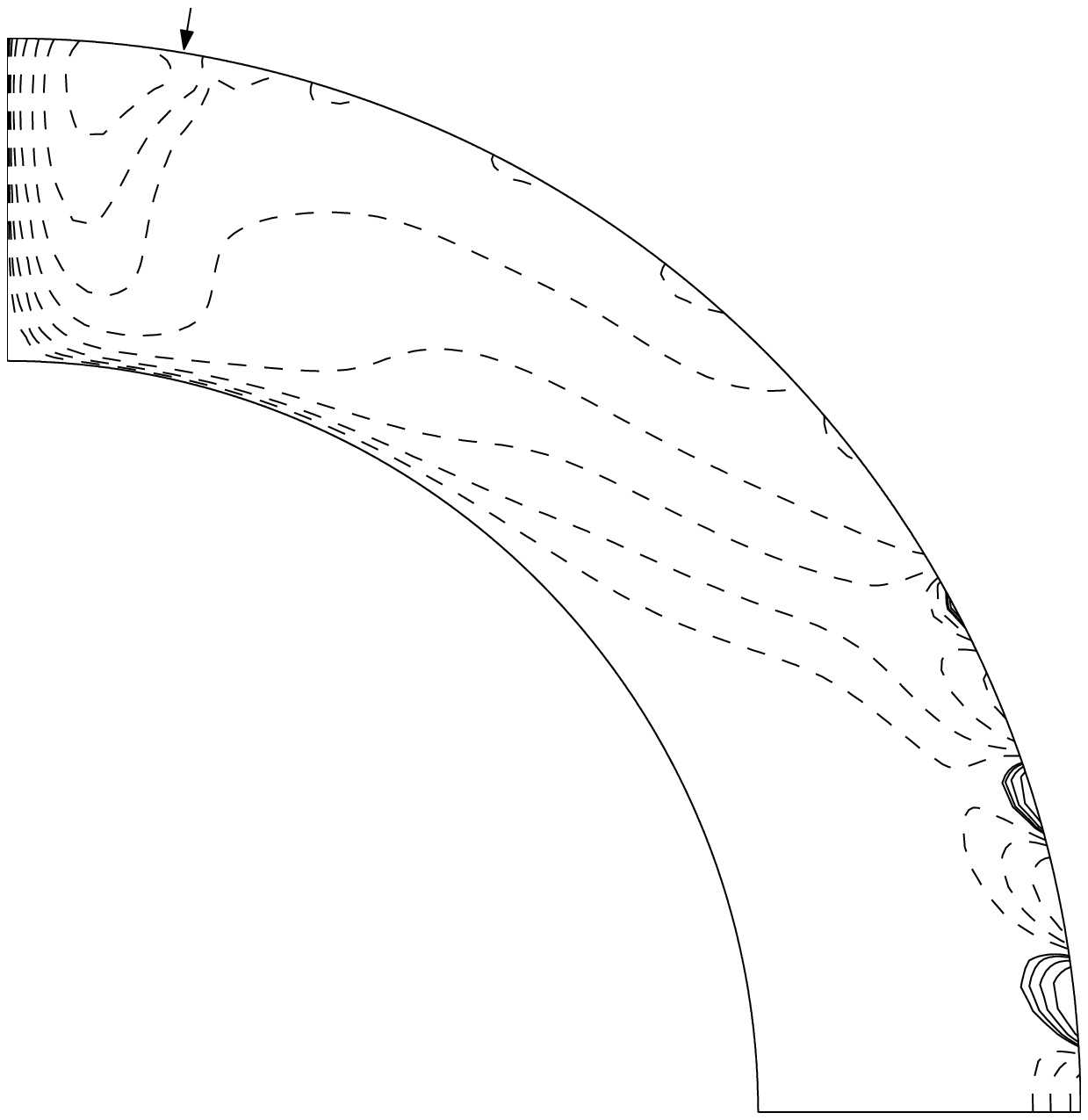,width=4cm,angle=0}
\vfill
\epsfig{file=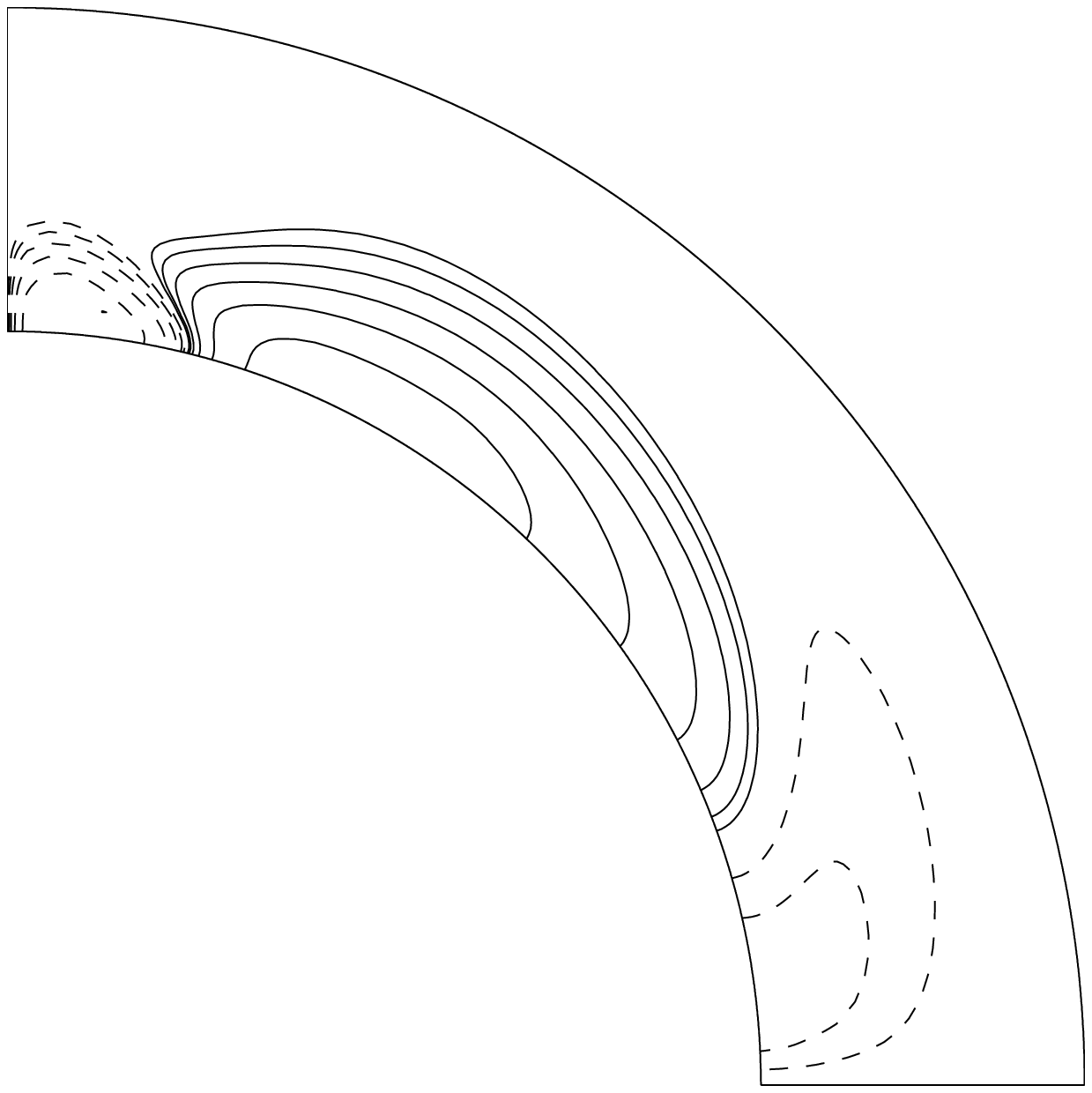,width=4cm,angle=0}
\epsfig{file=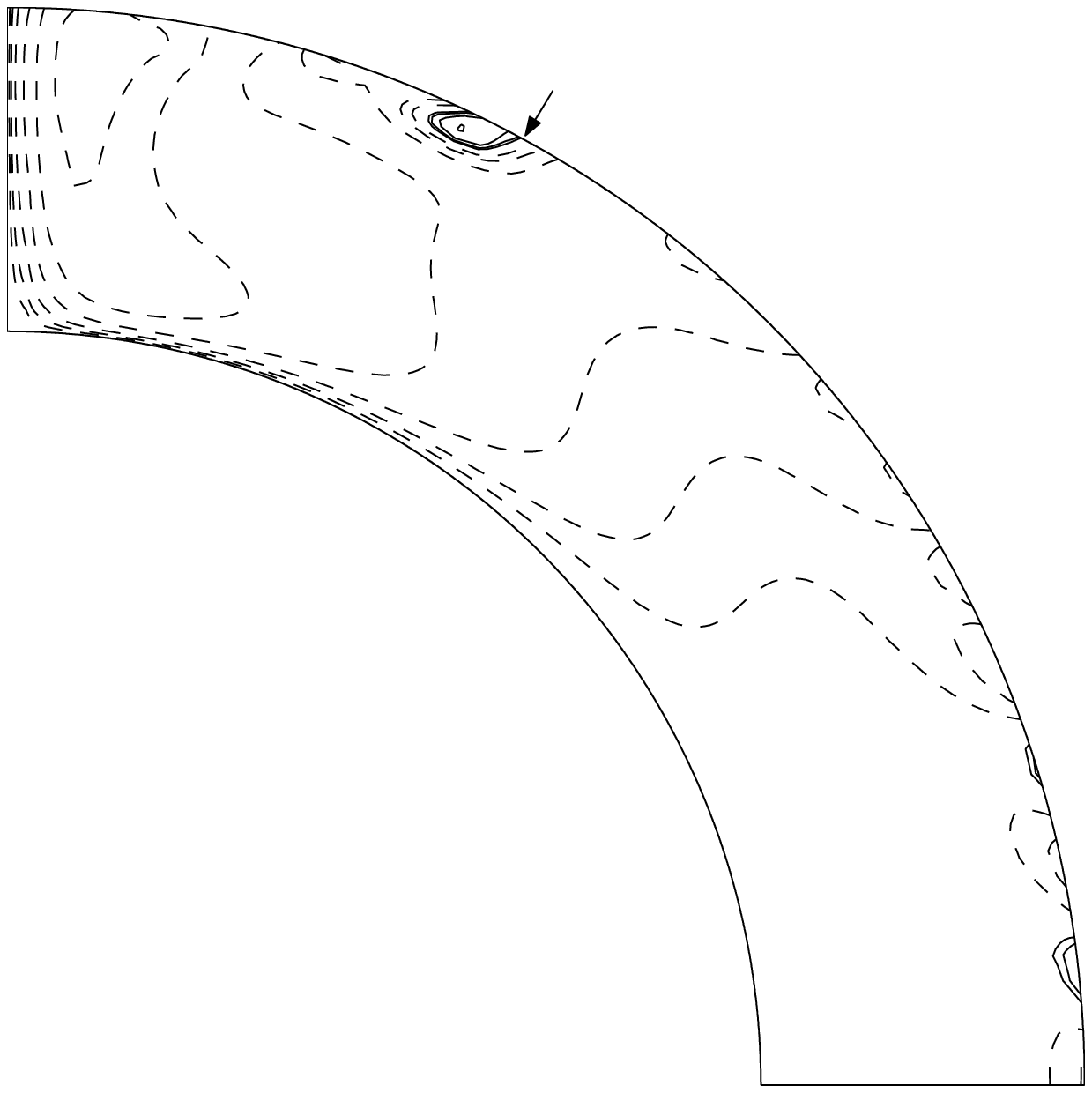,width=4cm,angle=0}
\vfill
\epsfig{file=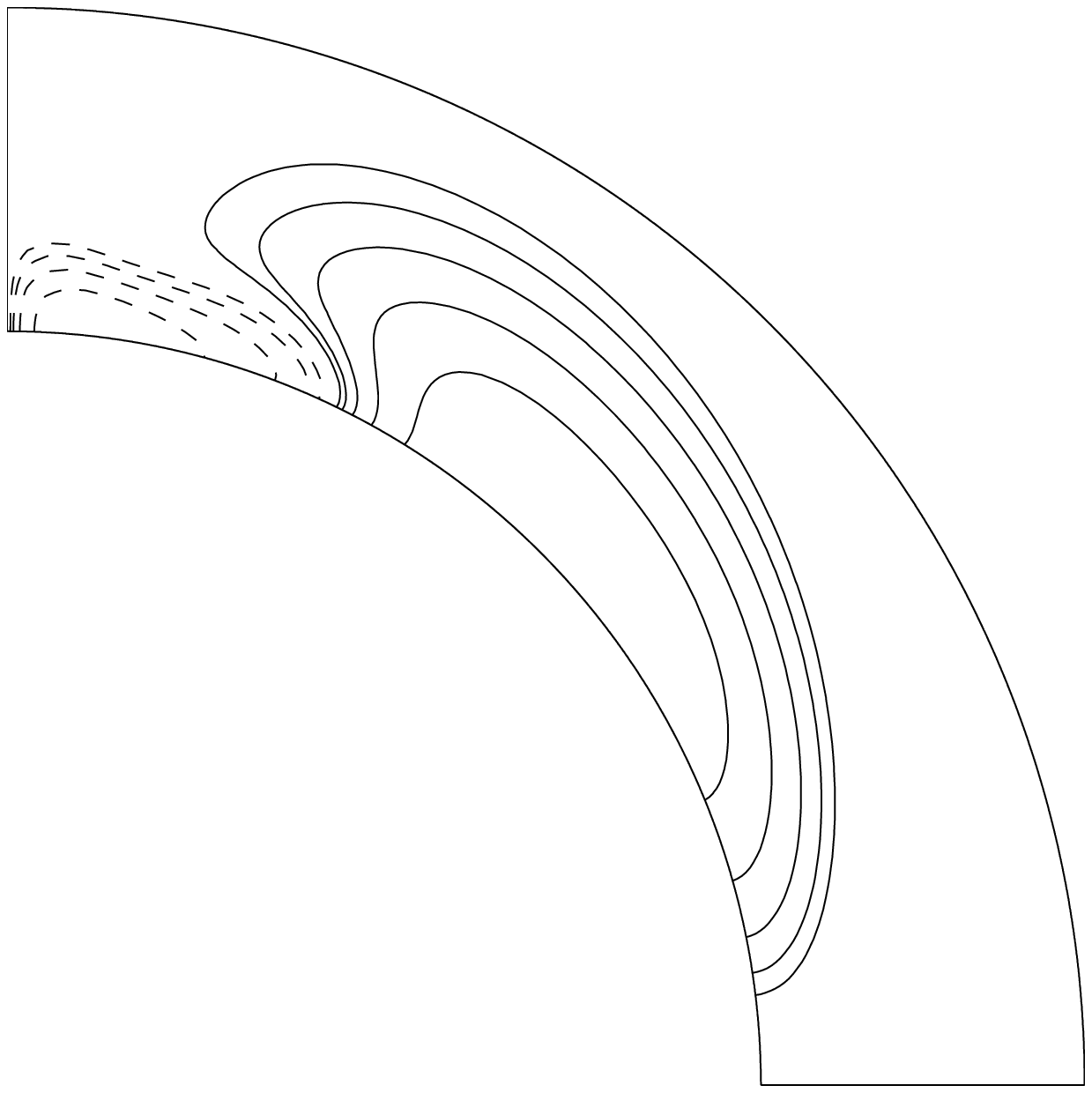,width=4cm,angle=0}
\epsfig{file=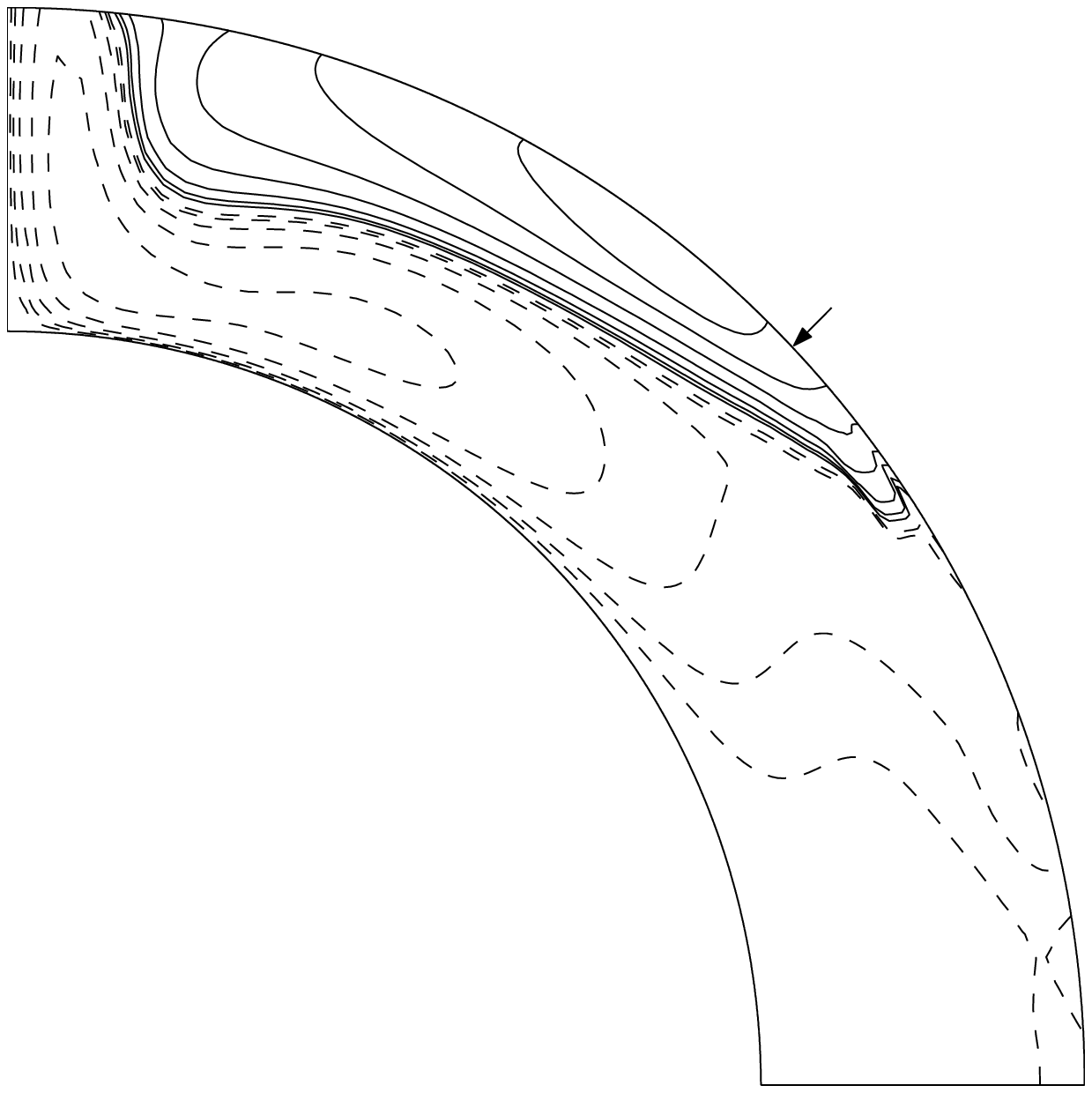,width=4cm,angle=0}
\vfill
\epsfig{file=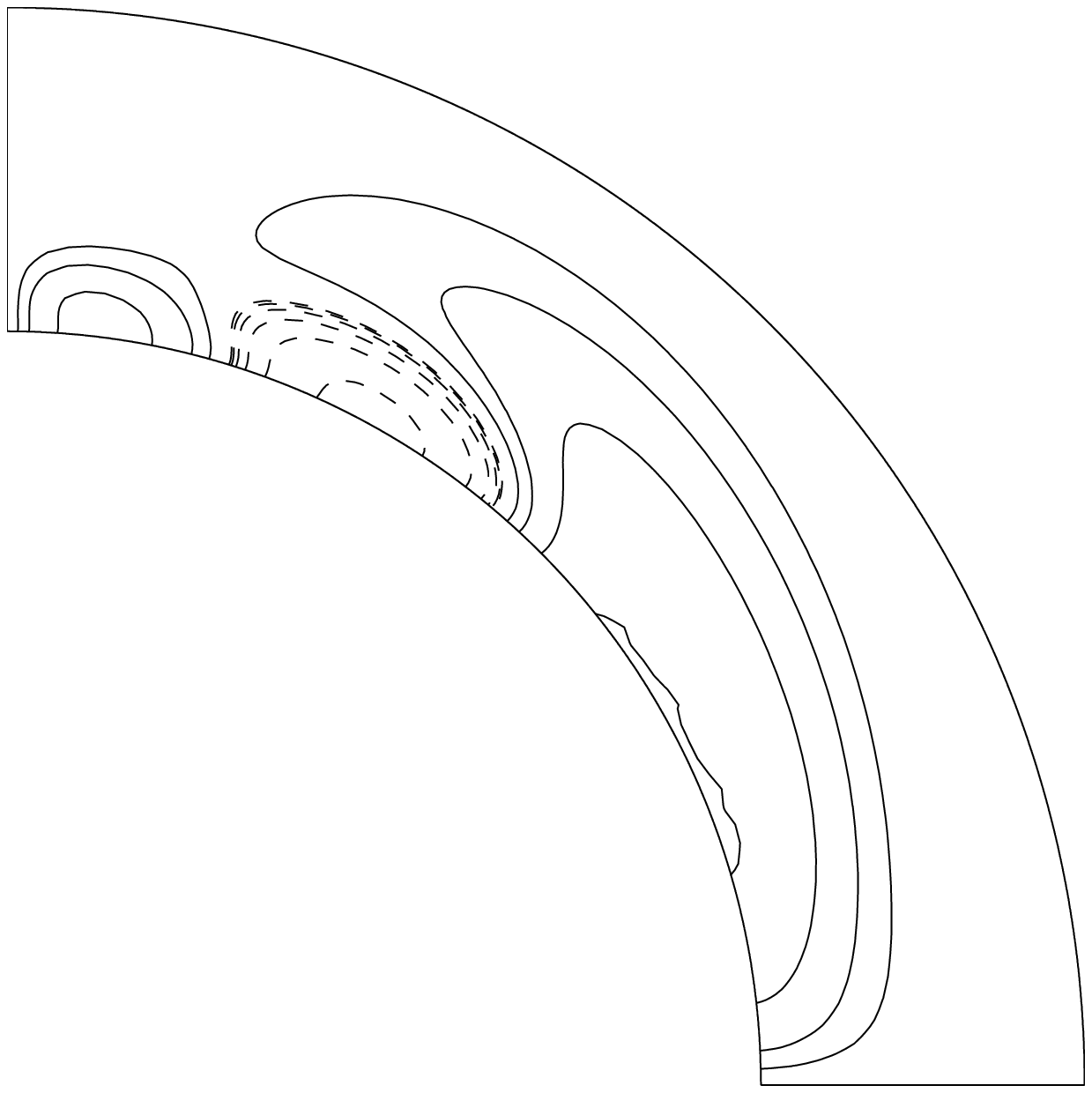,width=4cm,angle=0}
\epsfig{file=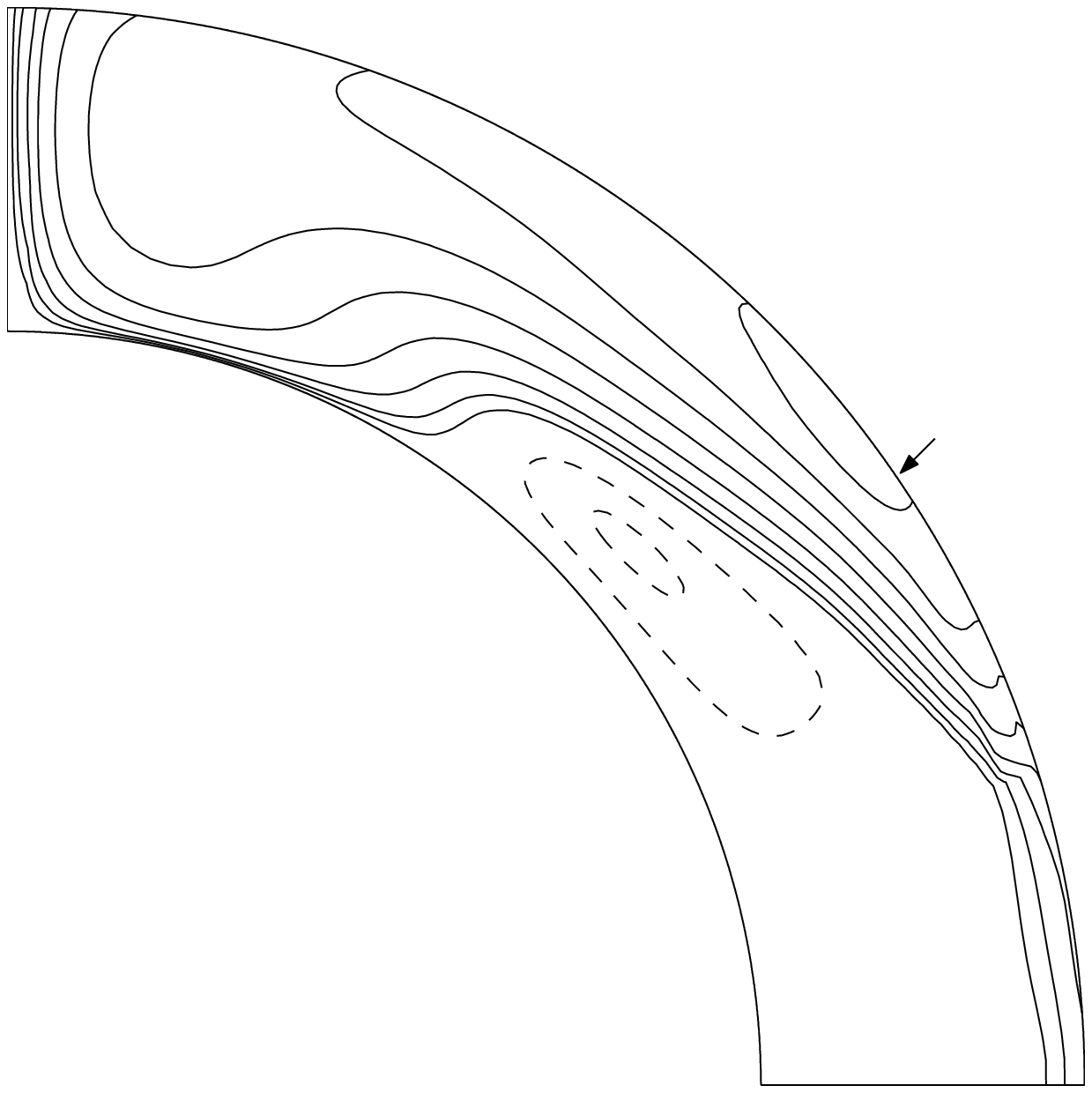,width=4cm,angle=0}
\caption{Time evolution of the toroidal field (left hand column)
and poloidal field (right hand column) configuration in a
meridional cut of the northern quadrant of the solar convection
zone ($0.7 \Ro, \leq r \leq \Ro$, $0 \leq \theta \leq \pi/2$)
for the case with $K' = 1000$, $f_d = 0.5$. The whole set covers 
a dynamo half period. That is from top to bottom $t$ = 0, $T_d$/8,
$T_d$/4, 3$T_d$/8, $T_d$/2.}   
\end{figure}      

\begin{figure}[h]
\epsfig{file=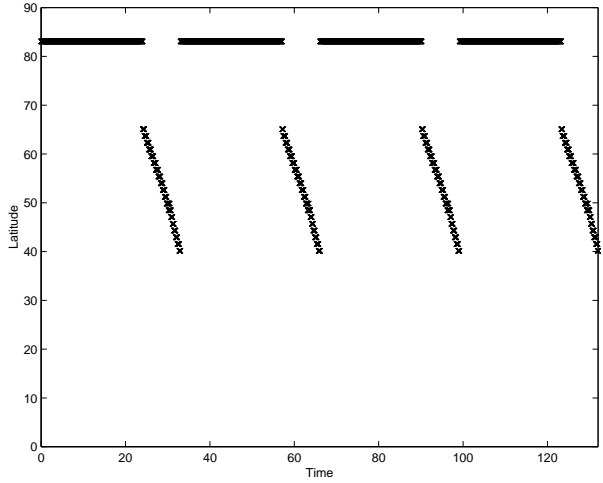,width=8.0cm,angle=0}
\caption{The variation in the eruption latitude with time (in years)
for $K' = 1000$, $f_d = 0$.}   
\end{figure} 

\begin{figure}[h]
\epsfig{file=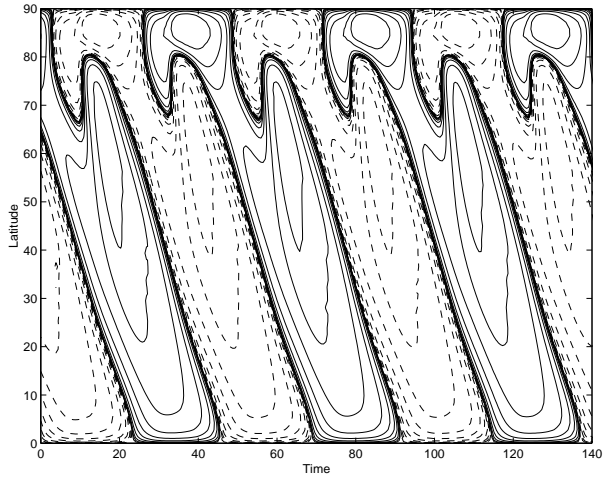,width=8.0cm,angle=0}
\caption{Time-latitude plot of the contours of constant toroidal
field $B$ at the bottom of the convection zone, for the case
with $K' = 1000$, $f_d = 0.5$ and ${v_0}=7.0 ms^{-1}$. The solid lines
denote positive $B$ and the dashed lines negative $B$. Time is
in years.}
\end{figure}

\begin{figure}[h]
\epsfig{file=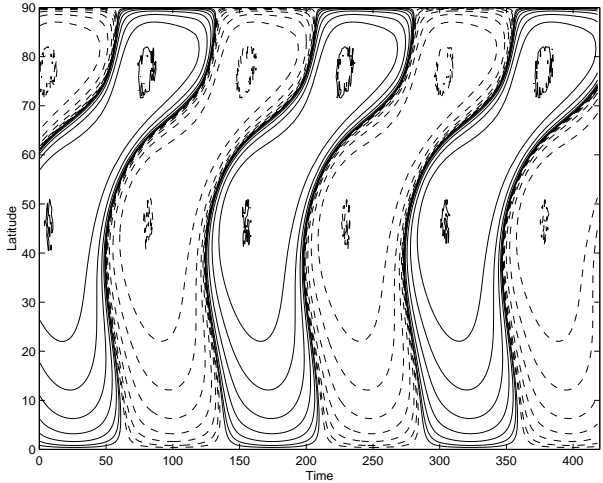,width=8.0cm,angle=0}
\caption{Time-latitude plot of the contours of constant toroidal
field $B$ at the bottom of the convection zone, for the case
with ${v_0}=0.0$. The other parameter are the same as in 
Figure~5.}     
\end{figure}  

\begin{figure}[h]
\epsfig{file=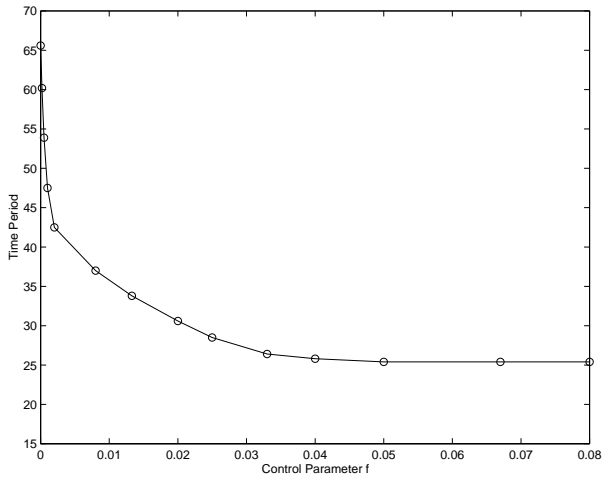,width=8.0cm,angle=0}
\caption{The variation in the dynamo period (in units of years)
with the control parameter $f$ for our second method -  
concentrated $\alpha$ effect with buoyancy.}   
\end{figure} 

\begin{figure}[h]
\epsfig{file=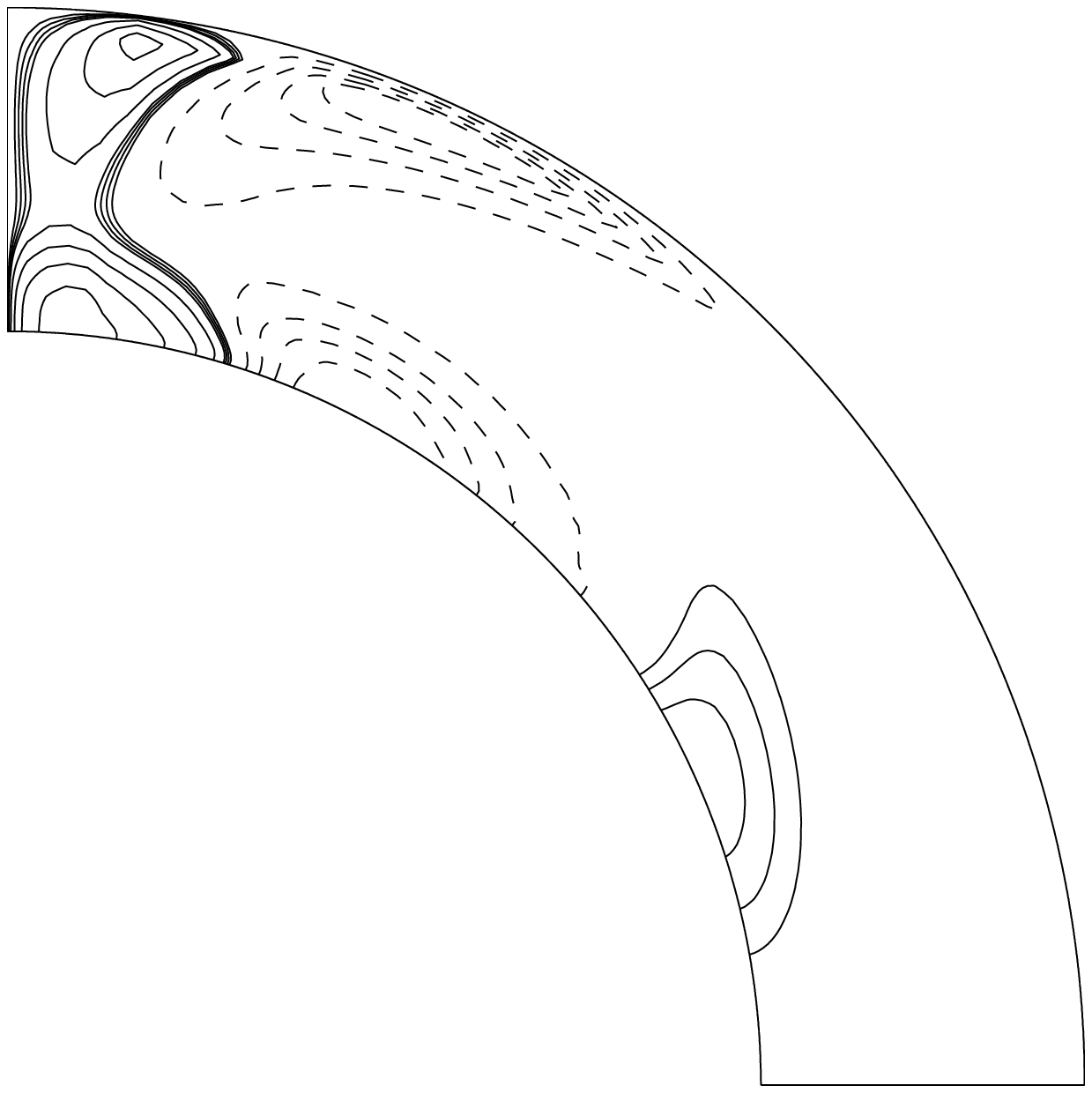,width=4cm,angle=0}
\epsfig{file=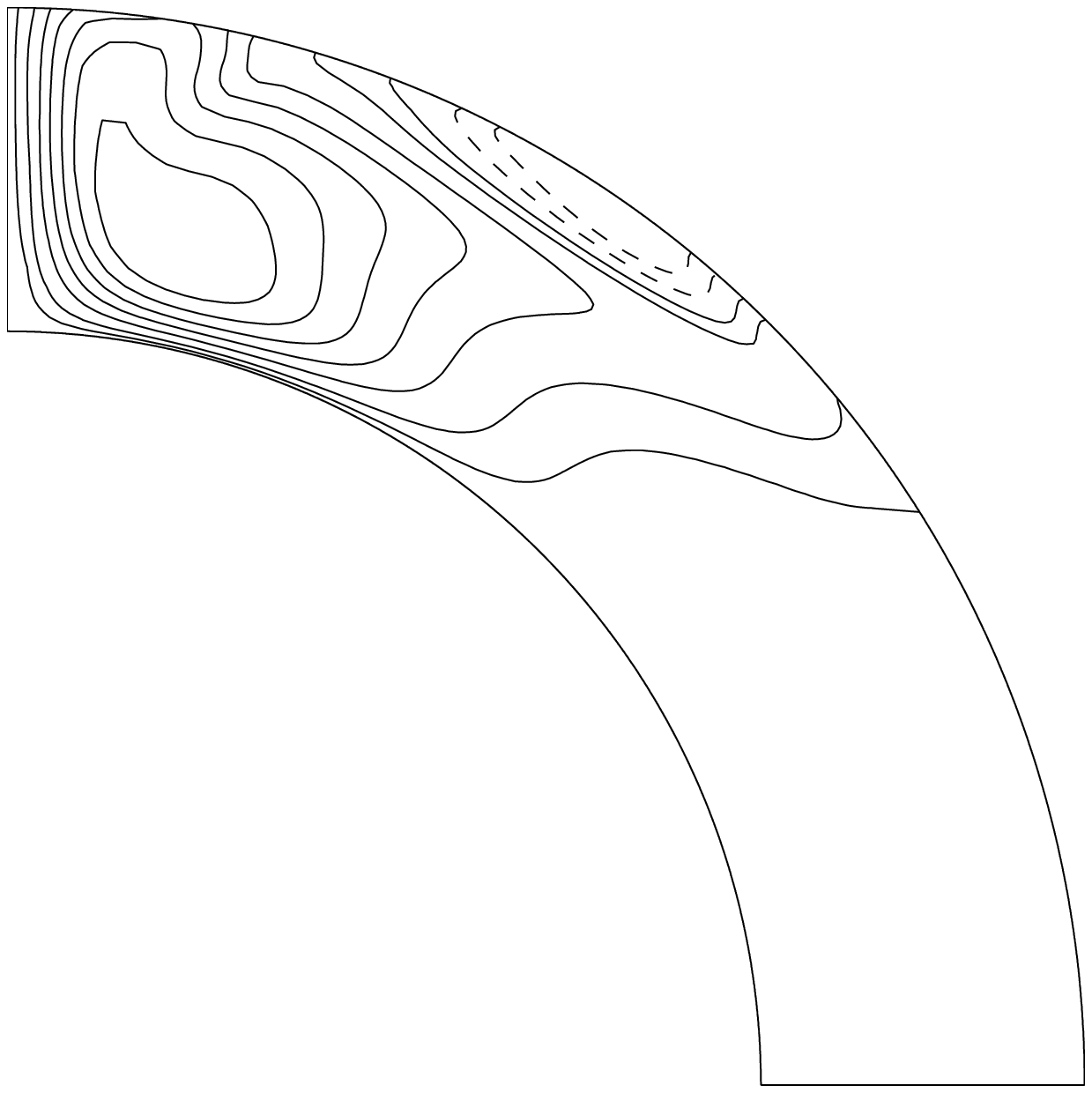,width=4cm,angle=0}
\vfill
\epsfig{file=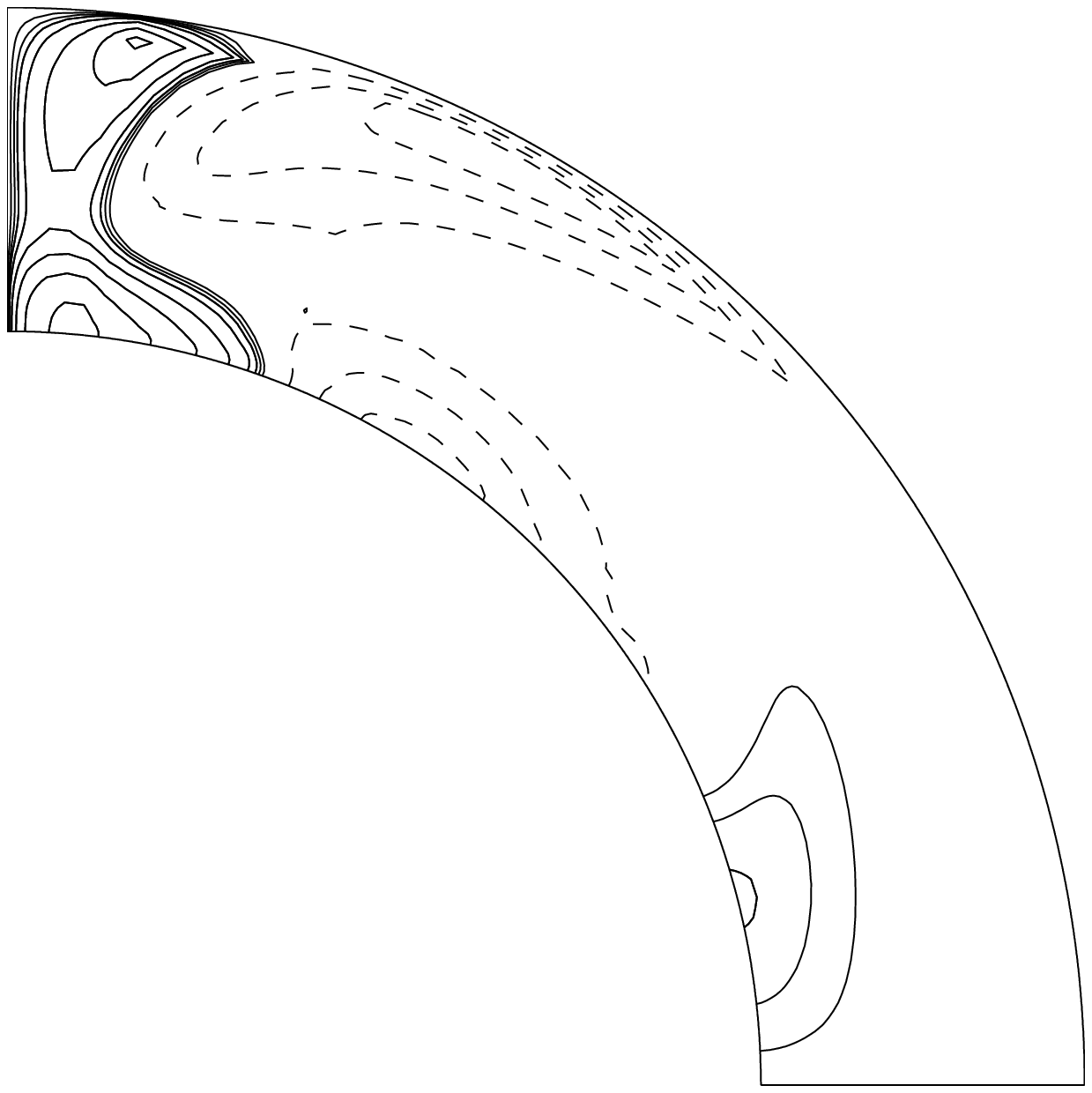,width=4cm,angle=0}
\epsfig{file=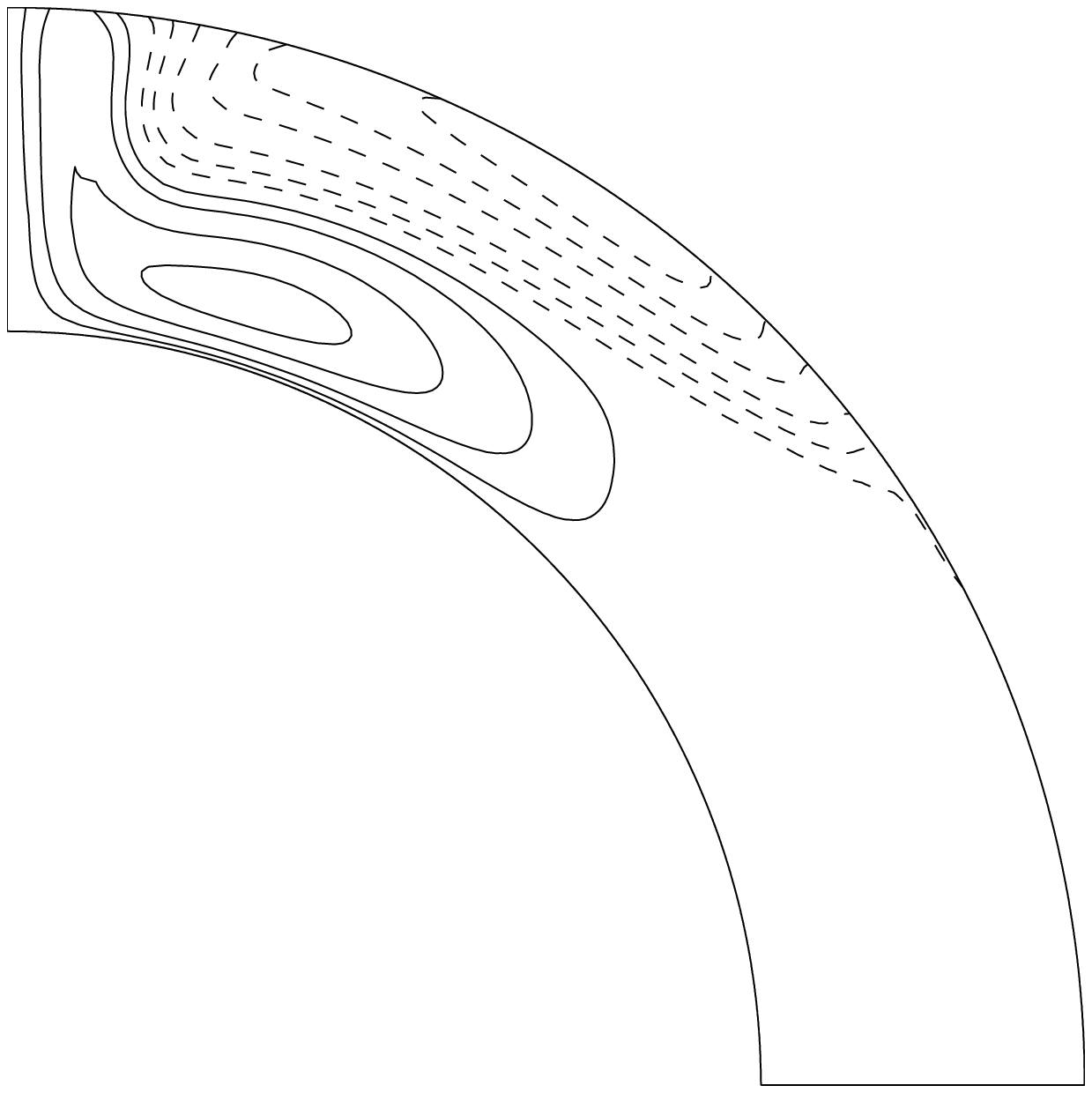,width=4cm,angle=0}
\vfill
\epsfig{file=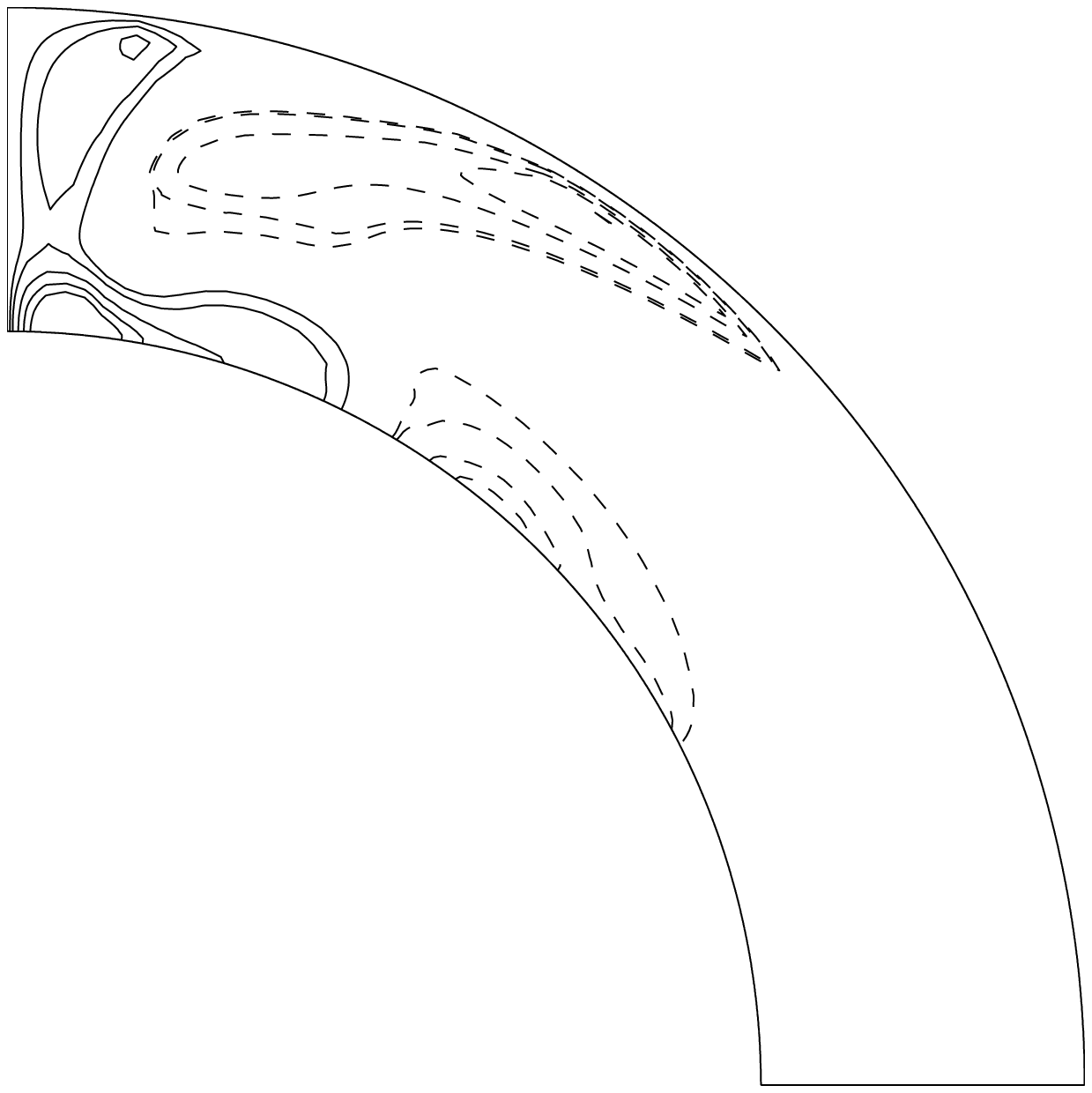,width=4cm,angle=0}
\epsfig{file=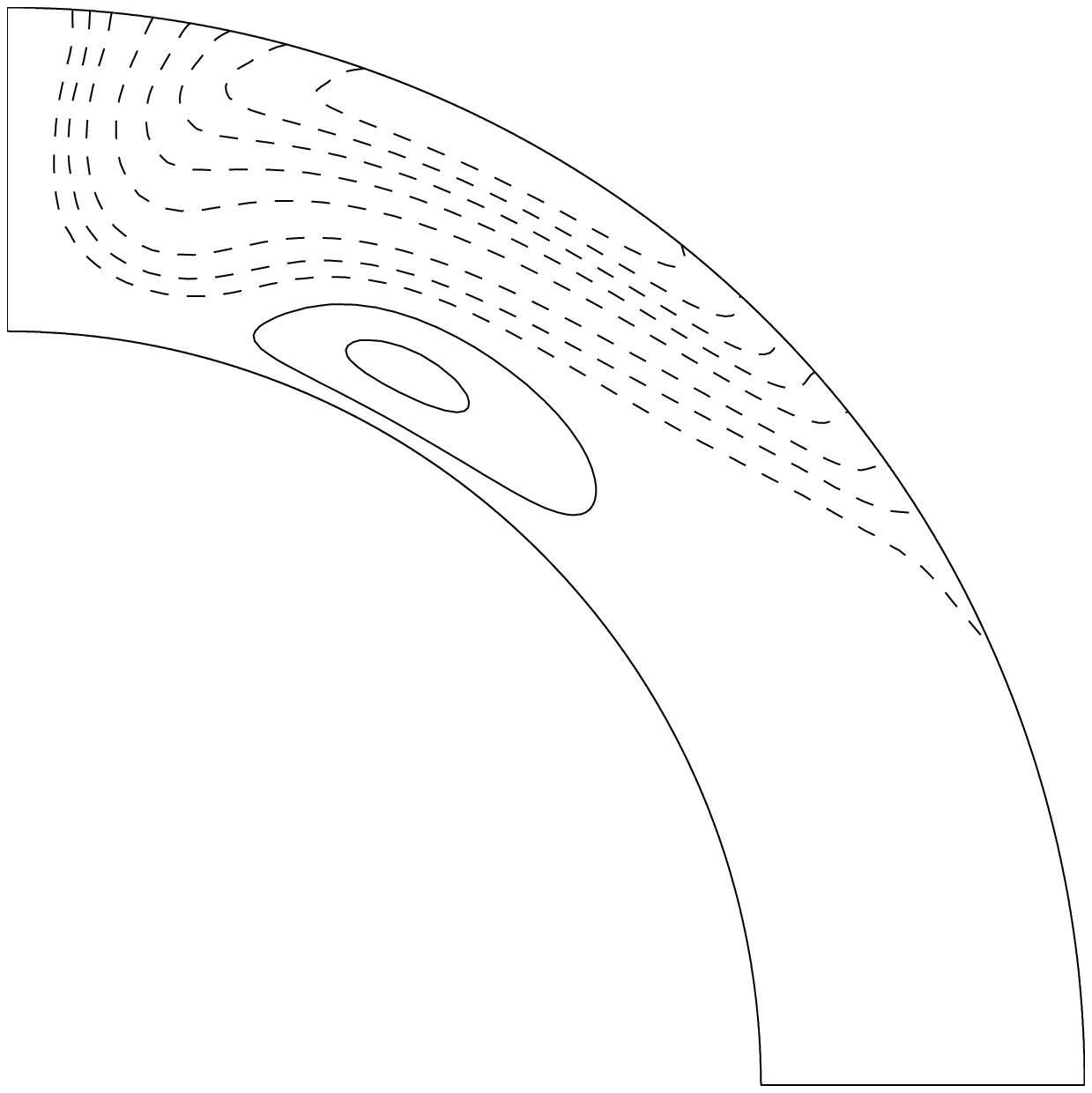,width=4cm,angle=0}
\vfill
\epsfig{file=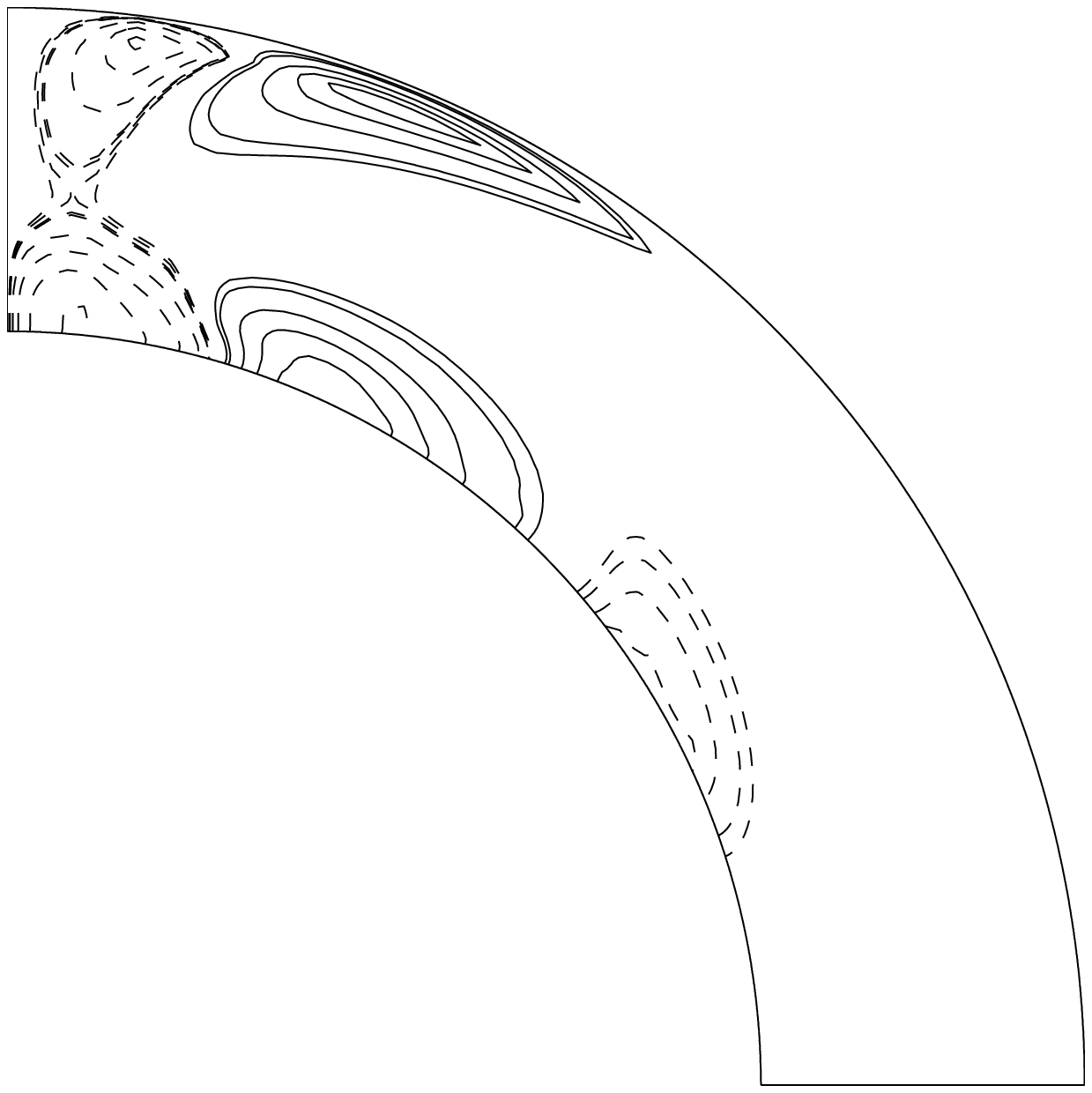,width=4cm,angle=0}
\epsfig{file=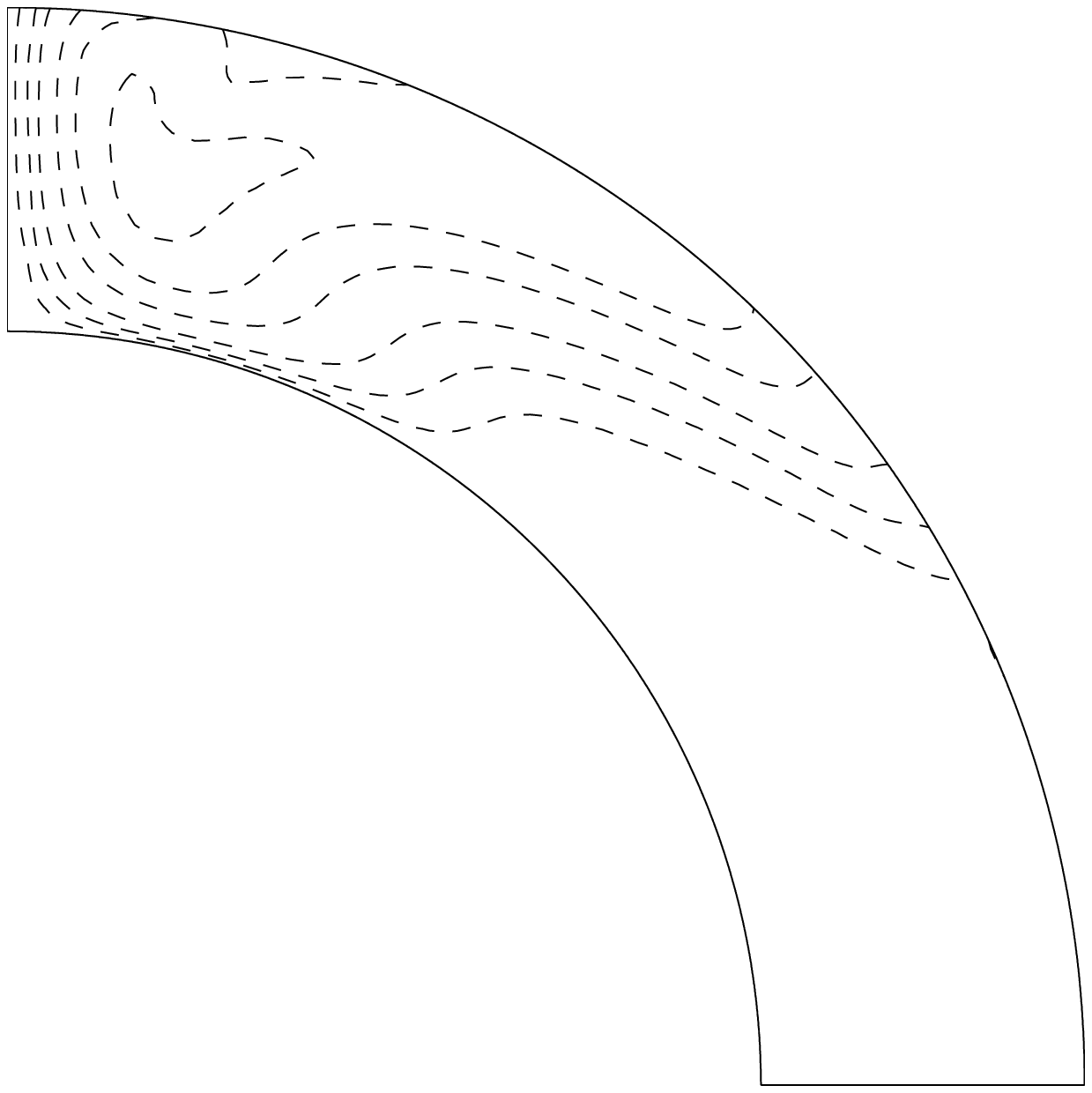,width=4cm,angle=0}
\vfill
\epsfig{file=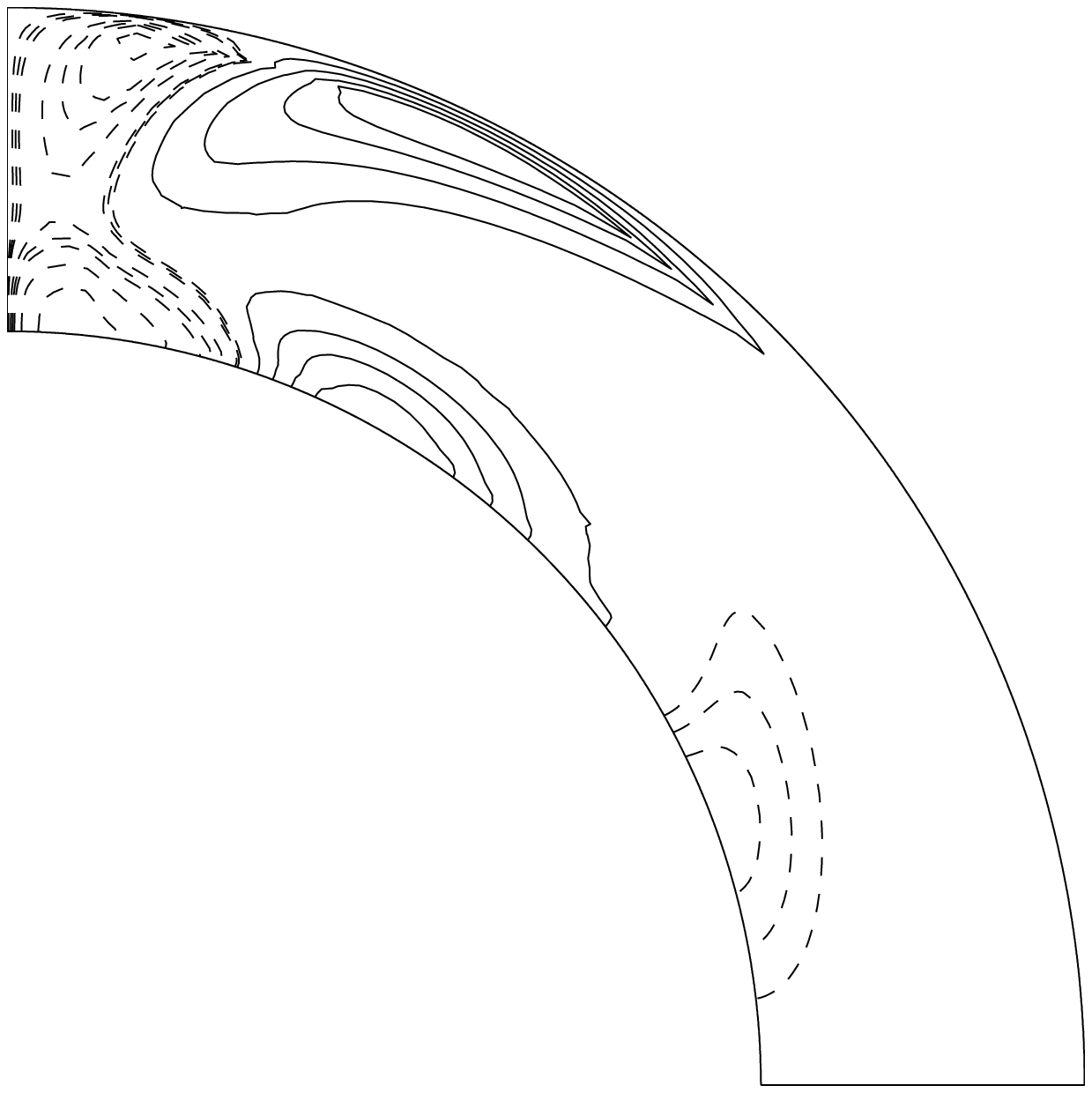,width=4cm,angle=0}
\epsfig{file=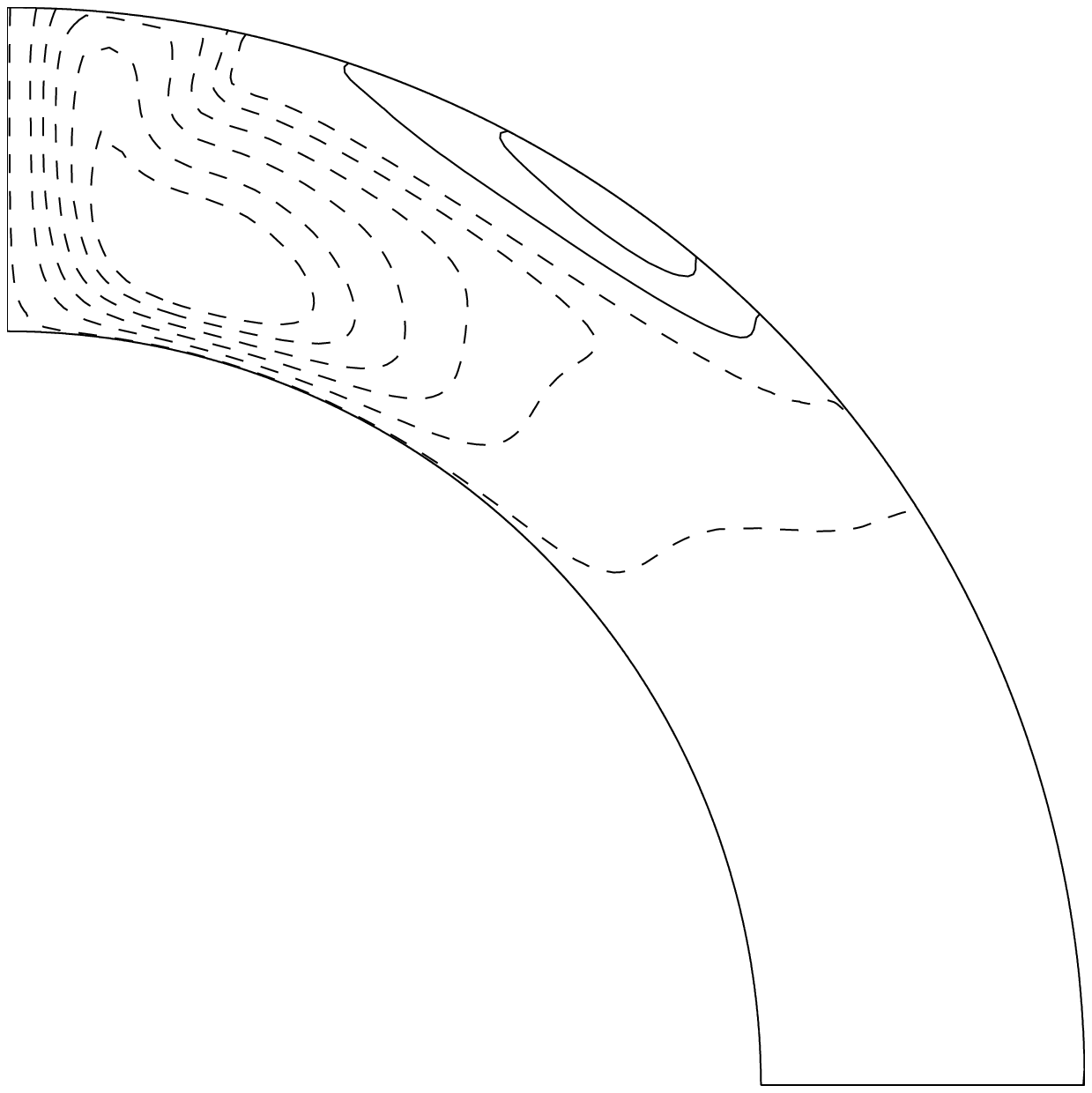,width=4cm,angle=0}
\caption{Time evolution of the toroidal field (left hand column)
and poloidal field (right hand column) configuration for the 
concentrated $\alpha$ effect with buoyancy method with $f =0.05$. 
The convention followed is the same as in Figure~3.}   
\end{figure}    

\begin{figure}[h]
\epsfig{file=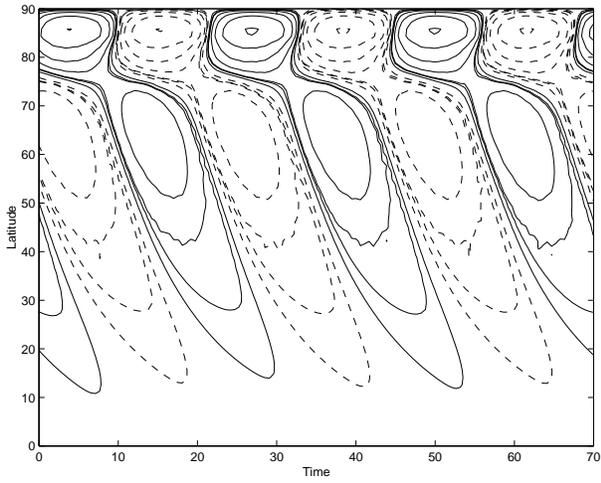,width=8.0cm,angle=0}
\caption{Time-latitude plot of the contours of constant toroidal
field $B$ at the bottom of the convection zone for the concentrated 
$\alpha$ effect with buoyancy method, with $f =0.05$. Time is in years.}
\end{figure}

\end{document}